# Thermodynamics Interpretation of Electron Density and Temperature Description in the Solar Corona

Daniel B. Berdichevsky[1,2], Jenny M. Rodríguez Gómez[3], Luis E. Vieira[4], Allison Dal Lago[4]

[1]College Park, Maryland, USA,[2] IFIR/UNR-CONICET, 2000 Rosario, Sta Fé, Argentina , [3] Skolkovo Institute of Science and Technology, Territory of innovation center Skolkovo, Moscow, Russia.[4]INPE, São José dos Campos, São Paulo, São Paulo, Brasil.

## Abstract

We reach a thermodynamic interpretation of CODET model and its prediction to electrons density and temperature grounded on the physics of hydro magnetism in global equilibrium. The thermodynamic interpretation finds consistency with the model with a magneto-matter medium that is diamagnetic, in the context of ideal magnetohydrodynamics (MHD). It is further noticed that the CODET predicts a polytropic anomalous index for the electron gas of the Sun's corona. It is shown that this unusual characteristic is consistent with assuming that the low quiescent solar corona is a magneto-matter state which possesses an underlying structure that was earlier described to explain the 2-D adsorption process by a surface of a solid of molecules of a gas at a given temperature and pressure by Langmuir. In our case, it is assumed that we are in the presence of a 3-D similar coalescence process, i.e. a Langmuir amorphous lattice in thermodynamic equilibrium. In this way, constitutive properties of the medium magnetic permeability, the non-dispersive acoustic speed, the expected equilibration time for the 1.1 to 1.3 $R_\odot$ , and energy density are determined quantitatively for most of the quiescent corona in a near solar minimum that extends for several months from 2008 to 2009.

## 1. Introduction

About half a century ago techniques to measure temperature and density in the Sun's corona were already known. Hence, electron density ($N_e$) and temperature ($T_e$) were studied using white light images from the corona (Van de Hulst, 1950, Newkirk et al, 1970, Guhathakurta and Fisher, 1995, Billings, 1995). However, there exist some uncertainties in these measurements, which are significant beyond 1 – 2$R_\odot$ above the limb (Habbal et al, 2013). This is a subject of great current interest. In addition, the conditions near the solar minimum of the low Sun corona provide to the heliospheric community a quantitative description of the solar corona during an extremely quiet long lasting time interval. In this regard, we hope to provide with this work a valuable contribution from the standpoint of a model empirical adjustment to data as well as a consistent theoretical interpretation of the model.

We propose to undertake the understanding of a relatively simple schema for predicting the temperature and density of electrons in the solar corona. This aspect of the solar corona we are discussing is what has been identified with the 'K-corona denomination.' For the non-expert, it is worth mentioning that the Sun's corona receives qualifiers that indicate the kind of its properties being addressed. We are interested in the Corona of the continuum (K-corona). This is the name used for the Sun corona when we consider the presence of a continuous emission spectrum. This spectrum is currently understood as being the photosphere light scattered toward us by the electrons gas in the corona (see e.g. Billings, 1996).

The quite simple description of $T_e$, $N_e$, shown in Figure 1, constitutes a global view of their evolution along the solar cycle, and a few other observables are derived using a method introduced recently, see Rodríguez Gómez et al, 2018. Specifically we concern ourselves here with the COronal DEnsity and Temperature (CODET) model's success in its qualitative description of the K-corona over more than 11 years, i.e. more than a solar cycle.

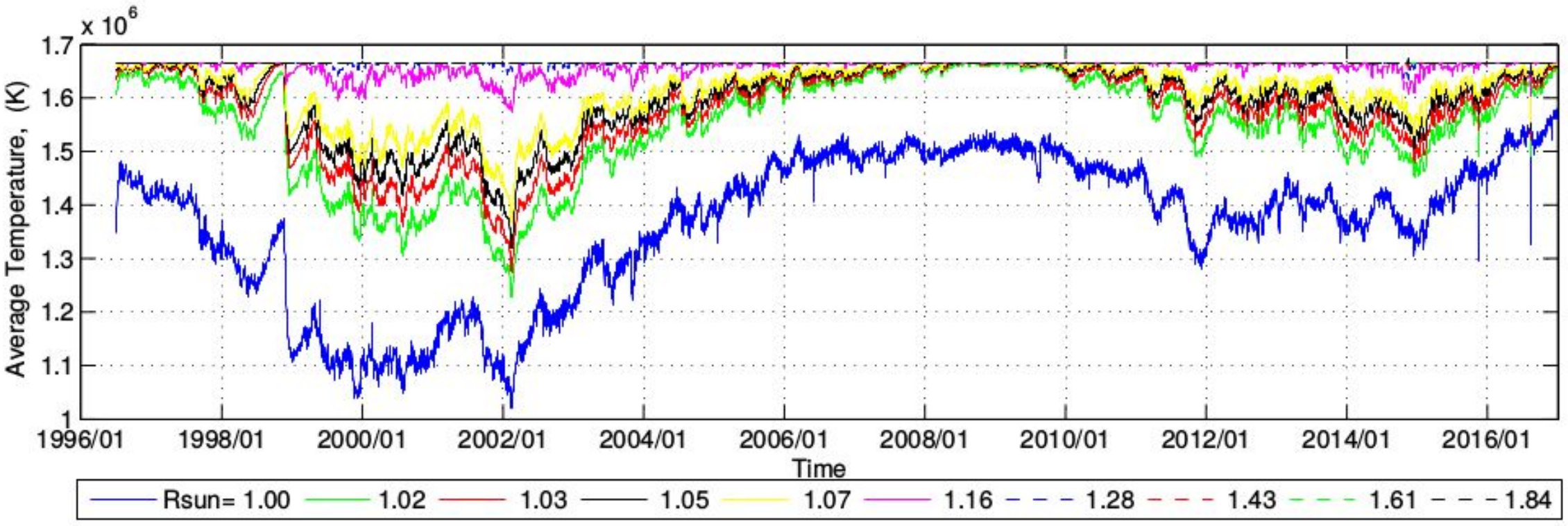

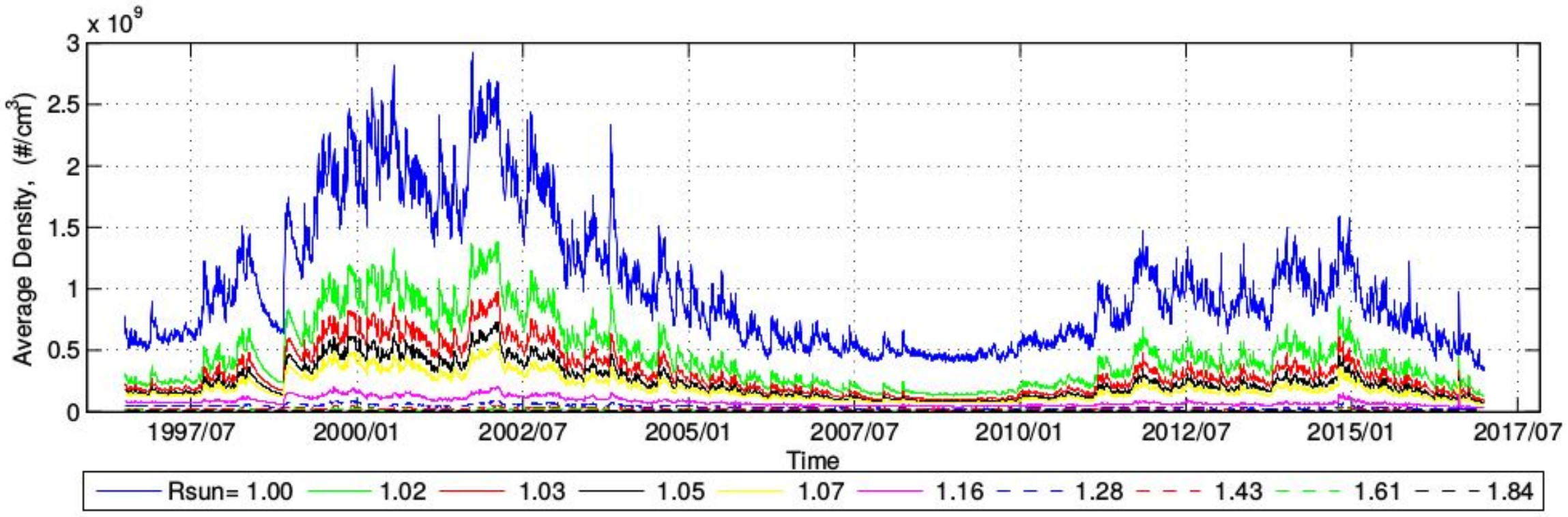

**Figure 1.** View of the low corona dependence on altitude of electrons $T_e$ (top) and $N_e$ (bottom) from CODET model and their progression in time covering cycle 23 and most of cycle 24.

These results illustrate that our study has the potential to be a beneficial contribution to the understanding of the Sun corona at least under particular conditions when we consider the basic questions raised in Withbroe, 1988 when stating that "The physical conditions in the Sun corona are vital to the development of an understanding of the mechanisms which heat the coronal plasma; but exists uncertainty of these mechanisms. Some models ignore the effects of the inward flow of energy carried by thermal conduction from hottest layers of the corona. They also ignore the effects of radiative losses in the low corona and chromosphere, corona, transition region." These statements, at least in part, sound true today. There are gaps, especially regarding a lack of understanding when it comes to the high temperature (more than $10^6$ K) of a Sun Corona region starting possibly after the transition region (TR) above/near the chromosphere height, and up in altitude to a few solar radii, away from which it is understood that there begins the convection of matter an magnetic field. What likely develops into what is known as a *slow* solar wind (SW) see e.g. Sanchez-Diaz et al, 2016, Vasquez et al, 2017, moving away from the Sun in its travel of near 100 AU or more until its encounter with the local interstellar medium (LISM, as it is explored in-situ, see e.g. Burlaga et al, 2013, Richardson et al 2017, Cummings et al, 2016.)

This state of affairs is possibly so despite the much learned on small-scales, as well as the solar network scale,: i) helioseismology, see e.g. Zhao, Kosovichev, Sekii, 2010; ii) nano-flares, see e.g. Klimchuk et al, 2006; iii) resolving some of the scales of spicules normal and type II, see e.g. De Pontieu et al, 2007; iv) plasma-jets discovery, e.g. Raouafi et al, 2016. Progress has been immense in our ability to collect solar data with a higher increase in spatial and temporal resolution, see e.g. the high speed/resolution movies of energetic processes identified in the polar magnetic coronal holes of the Sun at their boundaries, which could be connected to a study by Zurbuchen, Schwadron, Fisk, 1997 on what was identified by them as the motion of footpoints of magnetic field lines proposed by Fisk, 1996.

Summing up, it can be said that over the last decades there has been noticed progress on the physical processes that most likely determine the observed properties of the Sun atmosphere/corona and their outward propagation phenomena (SW, Coronal Mass Ejections, current/plasma sheet, and so on). However, these models are valid only if they can account for fundamental properties of the plasma, electron temperatures, the electron and ions densities, their flow speeds, as well as magnetic field strength and direction. Unfortunately, these measurements are not known that well in long periods of time like a solar cycle.

Although, $N_e$ and $T_e$ can be inferred from remote sensing images of the inner corona, and in-situ measurements in the interplanetary medium. When these models are effective, e.g. Rodríguez Gómez, 2017, require further understanding of the underlying physics. In this work we attempt to fill the gap for the simplest case scenario, i.e. the quiescent solar corona.

Furthermore, we notice that a recent work by Morgan and Taroyan, 2017 uses a similar global irradiance analysis of the Sun corona through the solar cycle from year 2010 to year 2017 than the present approach (Rodríguez Gómez, 2017, Rodríguez Gómez et al, 2018). This approach uses both the Differential Emission Model (DEM) and the scaling laws which describe the Temperature in the Solar Corona. However, in the work by Rodríguez Gómez covers a more extensive time interval (1995 – 2017) starting from the same basis of the magnetic field on the photosphere and employing comparable techniques to generate mean global changes, but it differentiates by further giving estimates of electrons density and temperature assuming a Corona in hydromagnetic equilibrium. Figure 1 is an example that Rodríguez Gómez, 2017 allows to recover variations in a large time scale.

Then, the extension in time of the quiescent conditions measured in our approach suggests us that the consideration of thermal equilibrium can be physically valid. We proceed to exploit in this work this assumption so as to infer in a consistent, ensemble based/statistical mechanics approach a few constitutive properties of the plasma in addition to the ‘CODET model’ provided temperature and number per cubic-cm of plasma particles, which possibly equals to approximately 3/2 to 9/5 the electrons number, when we consider that it is reasonable to expect that more of 80% of the ions likely are protons. In this sense we concentrate on the properties of the K-corona. In Section 2, we describe the CODET model and the main results for a solar cycle interval near its minimum (one year). These are the electron (*e*)-number and temperature profiles in three specific layers through the solar corona in a region limited between 1.1 and 1.3 $R_{\odot}$. Also, a description of the magnetic energy variation during this period is shown. This CODET model is designed for a quantitative description of the corona and valid for solar minimum, the long lasting solar minimum starting in 2008 and extending in time until the end of the year 2009. In Section 3 we combine remote solar corona observations from a variety of instruments in SOHO, details from the total solar eclipse from August 1, 2008 and earlier solar eclipse studies, and in-situ SW observations lasting several Carrington rotations to constraint the possibilities of the kind of solar corona to which we will apply our assumptions of its thermal properties, as they are derived from the global model (CODET) which provide the very good qualitative description as seen in Figure 1, panels 1, and 2 for an interval that extends from the solar minimum which connect solar cycles 22 and 23 up to passing the solar maximum of solar cycle 24. Also, the magnetic and kinetic pressure in this period, indication of the presence of ARs are further explored in the corona with the help of EUV images from EIT/SOHO. We add a consideration that allows us to neglect the impact of equatorial coronal holes (CHs) present during extended intervals of the solar minimum considered. In Section 4 we present two thermodynamic interpretations of the CODET model, testing its ability to perform quantitatively $T_e$, and $N_e$ predictions for the low corona. The thermodynamic interpretations further provide insight through their implications about the plasma properties in the region of the Sun corona. Section 5 touches on the important subject of the quiescent solar corona temperature conditions, brought to attention by our *steady-state magneto–matter* interpretation of Section 4. Section 5 further assesses the limits of the approach and reviews briefly some current lead candidates to an explanation of the high temperature of the Sun corona, as well as their lack of conclusive observational support. Discussion and conclusions are drawn in Section 6.

## 2. The COronal DEnsity and Temperature (CODET) Model

The CODET model uses magnetic field data provided by MDI/SOHO (Scherrer et al. 1995) and HMI/SDO (Scherrer et al. 2012). The Potential Field Source Surface (PFSS) was used to obtain the structure of the coronal magnetic field from 1 to 2.5 $R_{\odot}$ (Schrijver 2001; Schrijver & De Rosa 2003). We used a model based on Chianti atomic database 8.0 (Del Zanna et al. 2015). The emission model describes the characteristics of the EUV produced lines; 17.1 nm (FeIX), 19.3 nm (FeXII), and 21.1 nm (FeXIV). The optimization algorithm Pikaia (Charbonneau 1995) was used to research for the best fit-parameters, which describe adequately the density and temperature profiles through the solar atmosphere (Figure 1).

The optimization algorithm plays an important role in the CODET model because it is key to the connection between observed and modeled irradiance (Figure 2). Some tests were made using both individual and multiple wavelengths. The best results for the present project were obtained using three wavelengths simultaneously.

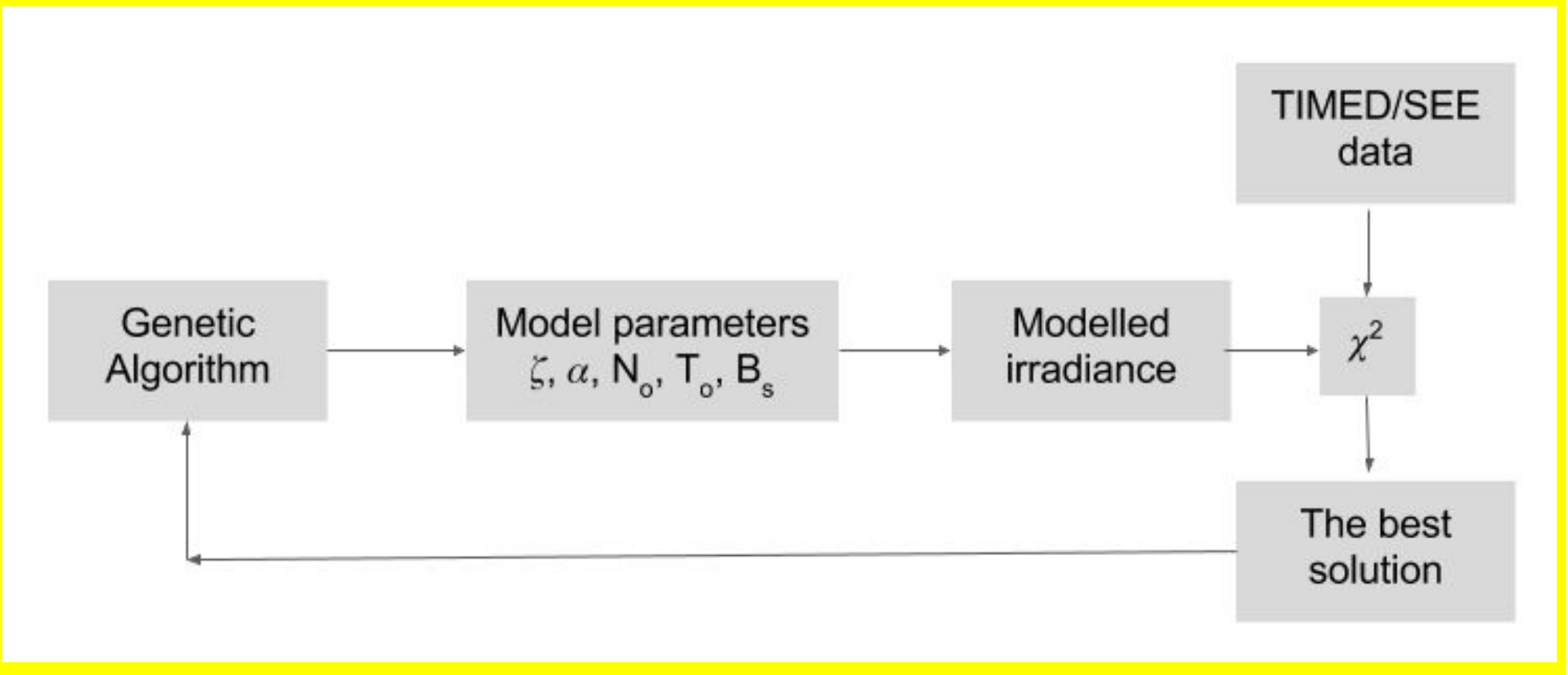

**Figure 2.** Flow diagram of the process involved in the generation of the CODET model as developed recently, see Rodríguez Gómez, 2017.

To obtain the goodness-of-fit between modeled and TIMED/SEE data we proceeded to perform the Chi-square test described below. $\chi^2$ function was defined as

$$\chi^2 = (I_{model} - I_{obs})^2 / |I_{obs}| \tag{1.a}$$

where $I_{obs}$ corresponds to the irradiance measured by TIMED/SEE, and $I_{model}$ is the modeled irradiance in all wavelength (17.1nm , 19.3nm and 21.1nm). The goodness of the fit is defined by

$$\chi^2 = \chi^2_{17.1nm} + \chi^2_{19.3nm} + \chi^2_{21.1nm} \tag{1.b}$$

In this case the goodness of fit is for $\chi^2 = 0.0010$. This value was reached with the model parameters presented in Table 1. It should be noted that our fitting scheme departs from the standard $\chi^2$ *-test* definition. These approach showed a far better model parameters optimization, than the more commonly found in the standard statistics books $\chi^2$-fit definition. These model parameters, listed also in Table 1, were then used to derive electrons gas temperature and number particles per $cm^3$ for the interval of interest, which is the quiescent Sun corona at extended solar minimum concentrating our analysis in the second part of year 2008.

In this approach, the scaling laws allow a description of the electron gas density and temperature profiles. For this purpose, we employed density and temperature as a function of the magnetic field (see e.g. Robbrecht et al, 2010; Yokoyama and Shibata, 2001; Golub, 1983; Emslie, 1985). On this occasion, an extremely Quiet Sun corona, the description does not use the loop length of each magnetic field line. We show the main results related to density, temperature, magnetic energy, magnetic and kinetic pressure. A small interval from Jan 1, 2008 to Jan 1, 2009 is considered. Figures 3 and 4 show, for these quiescent solar conditions, a year's description of the electron gas temperature and density (see Figure 3) dependence with altitude to the photosphere. The model parameters that are used are listed in Table 1, and the scaling law relationships, including equations and illustrative Figures showing changes as a function of height are outlined in the following subsection 2.1 to 2.4.

**Table 1.** Constants and the parameters of the model

| Parameter | Value | Instrument/ Dataset | Model | Reference |
|---|---|---|---|---|
| $B(r/R_{\odot})$ for $r > R_{\odot}$ | Model evaluation | SOHO/MDI, SDO/HMI | PFSS[1] input to SFTM[2] | Aschwanden/ Schatten/ Schrijver |
| $R_{\odot}$ | $7x10^5 km$ | | Standar value | Photosphere distance to Sun center |
| $R_{Bth0}$ | 1.2 $R_{\odot}$ | | CODET[3] | Rodríguez Gómez, 2017 |
| $B_{th0}$ | $3.77\ G$ | | CODET | Rodríguez Gómez, 2017 |
| $B_{sat}$ | $4.175\ G$ | | CODET | Rodríguez Gómez, 2017 |
| $\zeta$ | $1.252$ | | CODET, New parameters | This work |
| $N_o$ | $2.95x10^8 e/cm^3$ | | CODET, New parameters | This work |
| $\alpha$ | $-1.4938$ | | CODET, New parameters | This work |
| $T_o$ | $1.66x10^6 K$ | | CODET, New parameters | This work |
| $f_{\tau}$ (cadence) | $1/24hrs$ | | CODET | Rodríguez Gómez, 2017 |

### 2.1 Density profiles

The density profile (Figure 3) is described by the following expression:

[1] **PFSS** = Potential Field Source Surface model, see Altshuler and Newkirk, 1969
[2] **SFTM** = Surface Flux Transport Model of photosphere magnetism, developed by Schrijver, 2001
[3] **CODET** = Integrated model in Rodríguez Gómez, 2017. This model uses as input the extrapolated radial magnetic field through the solar atmosphere from the PFSS model, and also the magnetic flux transport model SFTM. It is shown that conservation of magnetic flux is assumed in the model

$$N = N_o (B/B_{sat})^{\zeta} \qquad (2)$$

where $B_{sat} = 4.17\ Gauss$, $N_o = 2.95x10^8$, and $\zeta = 1.252$, where $\zeta$ is the constant identified as $\gamma$ in Rodríguez Gómez, 2017, Rodríguez Gómez et al, 2018.

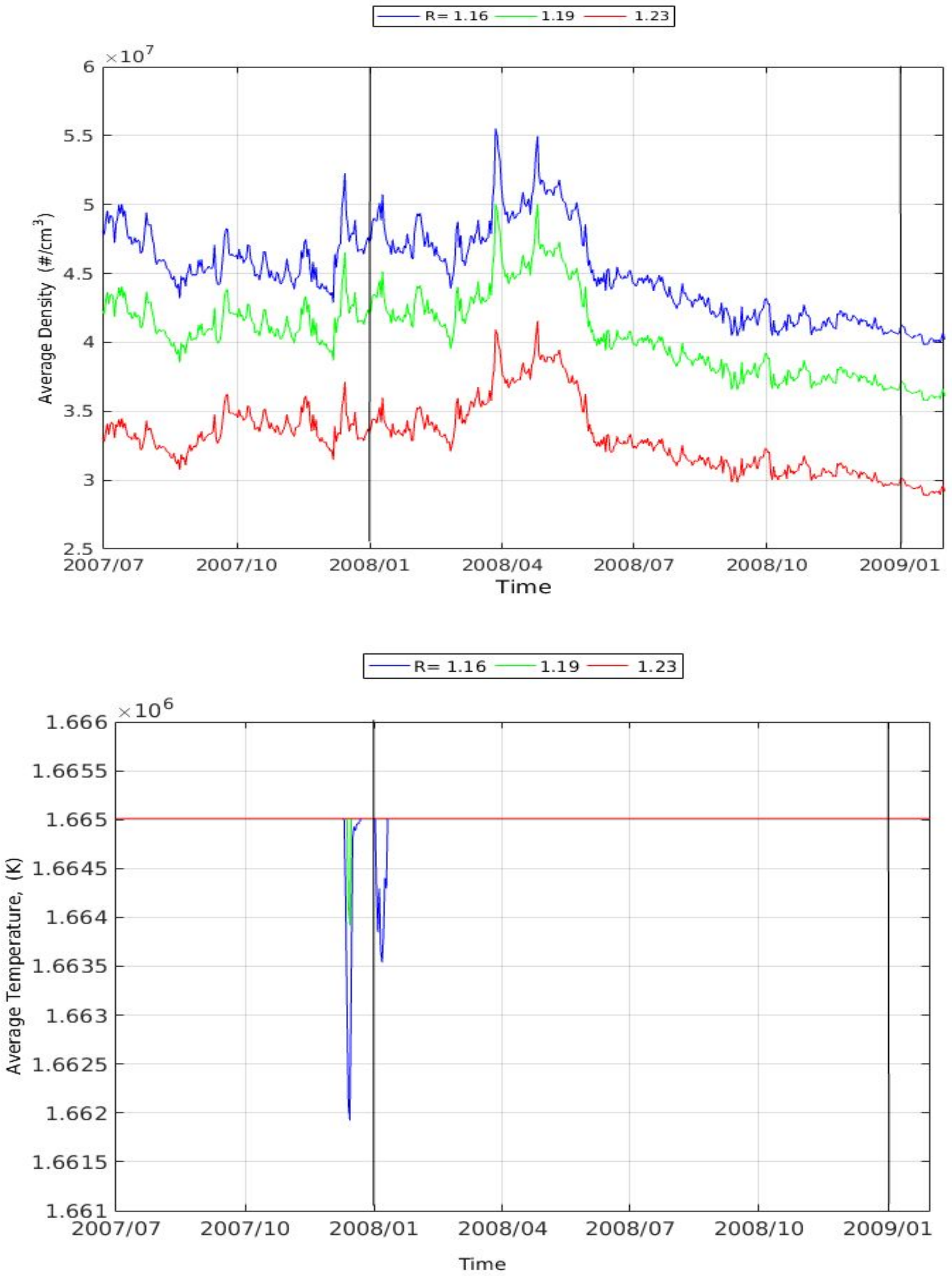

**Figure 3.** Density profile $N_e$, the number electrons per cubic cm (top panel) and Temperature profiles (bottom panel), the interval of interest in this work extends between thick vertical lines, from Jan. 01, (2008) to Jan. 01, (2009) at different heights: 1.16 (blue line), 1.19 (green line), 1.23 $R_{\odot}$ (red line).

The density profile shows an increase from 2008/03 to 2008/06 in all layers, as shown in Figure 3 left panel. After this period the density decreases in all layers.

The temperature profile definition is

$$B_{th} = B_{tho} e^{(-(r/Rbth)^2)} \quad (3)$$

where $B_{tho} = 20\ Gauss$ an $Rbth = 1.2\ R_\odot$. Using the following conditions, it is possible to define the temperature profile, see Figure 4:

if $B < Bth$ $T = T_0$ (4.a)

and if $B \geq Bth$ $T = T_0\,(B/Bsat)^\alpha$ (4.b)

where $T_0$, *Bsat,* and $\alpha$ used are listed in Table 1.

The right panel of figure 3 shows temperature profiles from Jan.01 (2008) to Jan.01 (2009) (black vertical lines demarcate the interval discussed in the work. Three different layers were considered R=1.16$R_\odot$ (blue line), R=1.19$R_\odot$ (green line) and R=1.23$R_\odot$ (red line). The temperature profiles show variation in 2008/01 (Figure 3, right panel). A decrease from 1.665x$10^6$ K (at R=1.19$R_\odot$ and R=1.23$R_\odot$*)* to 1.6636x$10^6$ K (at R=1.16$R_\odot$) was shown. The internal layer R=1.16$R_\odot$ show a lower value compared to R=1.19$R_\odot$ and R=1.23$R_\odot$ where the temperature profile is constant, due to the relation with the extrapolated magnetic field. It is important to stress that -after PFSS predictions- the extrapolated magnetic field is at each altitude near-constant in the three 'external' layers in the solar corona we consider.

## 2.2 Average magnetic energy density through the solar corona

The energy density is calculated using the following expression:

$$nB = energy/volume = 1/4\pi\ B^2/\mu_o \quad (5)$$

where $nB$ is the energy density, $B$ corresponds to the magnetic field 1.0 - 2.5$R_\odot$ (from PFSS) and $\mu_0$ is the magnetic permeability (4πx$10^{-7}$ kg m Amp$^2$/s$^2$) in the vacuum (it corresponds to an assumption in this model). The average energy density is obtained and from it the mean value in the solar atmosphere.

Figure 4 shows that the magnetic energy density through the solar corona shows an increase near to the 2008/04 and later this energy decrease shows small variations.

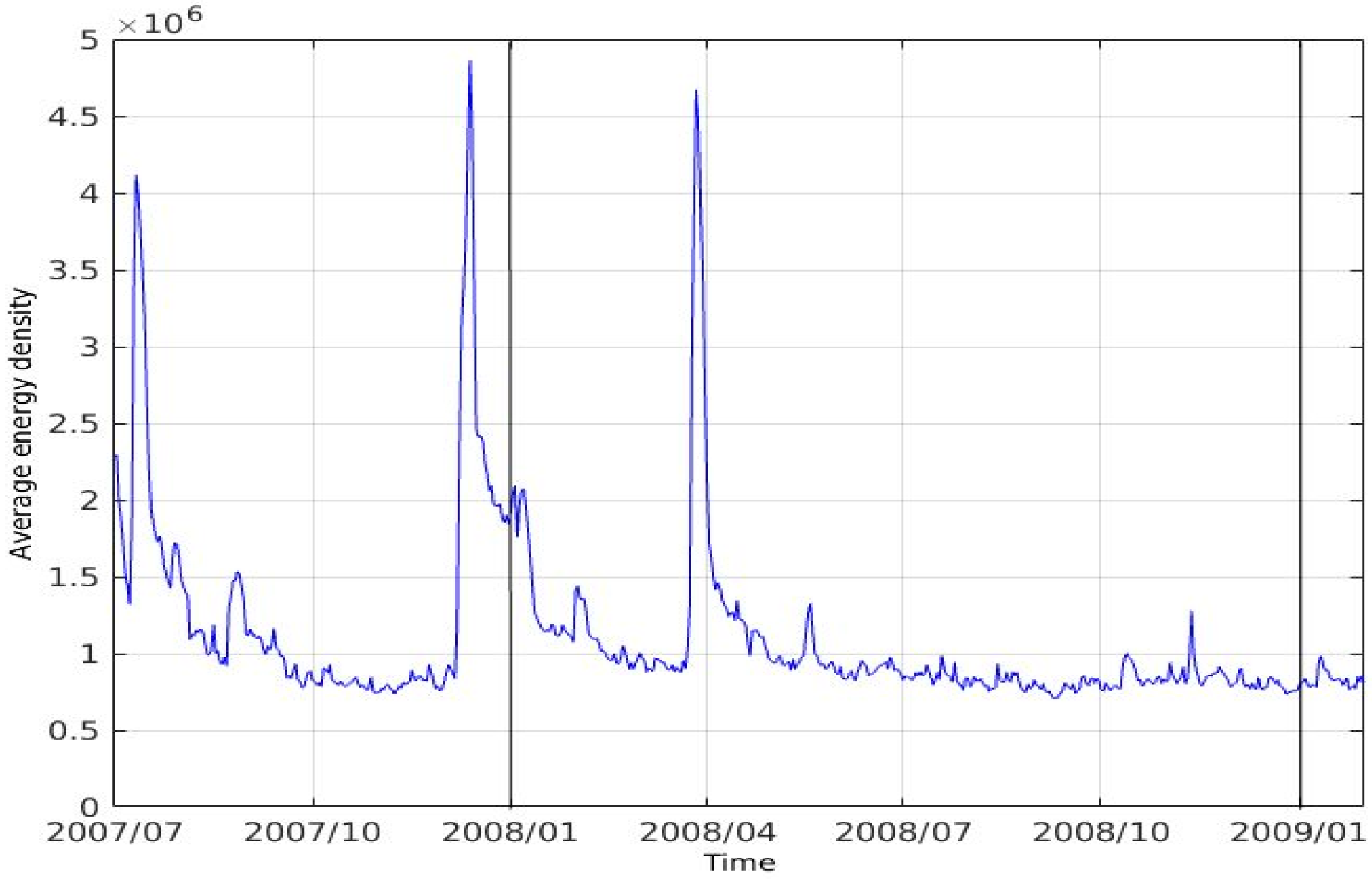

**Figure 4**. Average magnetic energy density, in ergs per cubic cm, vertical lines demarcate time interval.

## 2.4 Magnetic and kinetic plasma pressure

The magnetic energy pressure analysis was done using the magnetic field from PFSS

$$B = (B_r^2 + B_\varphi^2 + B_\theta^2)^{1/2} \tag{6}$$

where $B_r$ , $B_\varphi$ , $B_\theta$ are the magnetic field components from Potential Field Source Surface (PFSS) , see e.g. Altschuler and Newkirk, 1969. The magnetic pressure

$$P_m = B^2/(8\pi\mu_o) \tag{7.a}$$

Figure 5 (left panel) illustrates that the magnetic pressure shows increased values from 2008/03 to 2008/06. The maximum value is present in 2008/04, after that the magnetic pressure presents small variations related to a quiet corona.

The kinetic pressure analysis was made using the following expression

$$P = (N/Vol)k_B T \tag{7.b}$$

where N corresponds to the mean electron (e-) number and T is the e-temperature in each layer. $k_B$ is the Boltzmann constant and Vol is the volume we consider convenient to use in each layer. Notice that often observations give a direct estimate of the (N/Vol) either in remote or in-situ estimates (see values in Tables 1, and 2).

In general terms we can express

$Vol = 4\pi r^2 \delta r$

where '*r*' correspond to the radius in each layer of interest r =1.16, 1.19 and 1.23 $R_{\odot}$ .

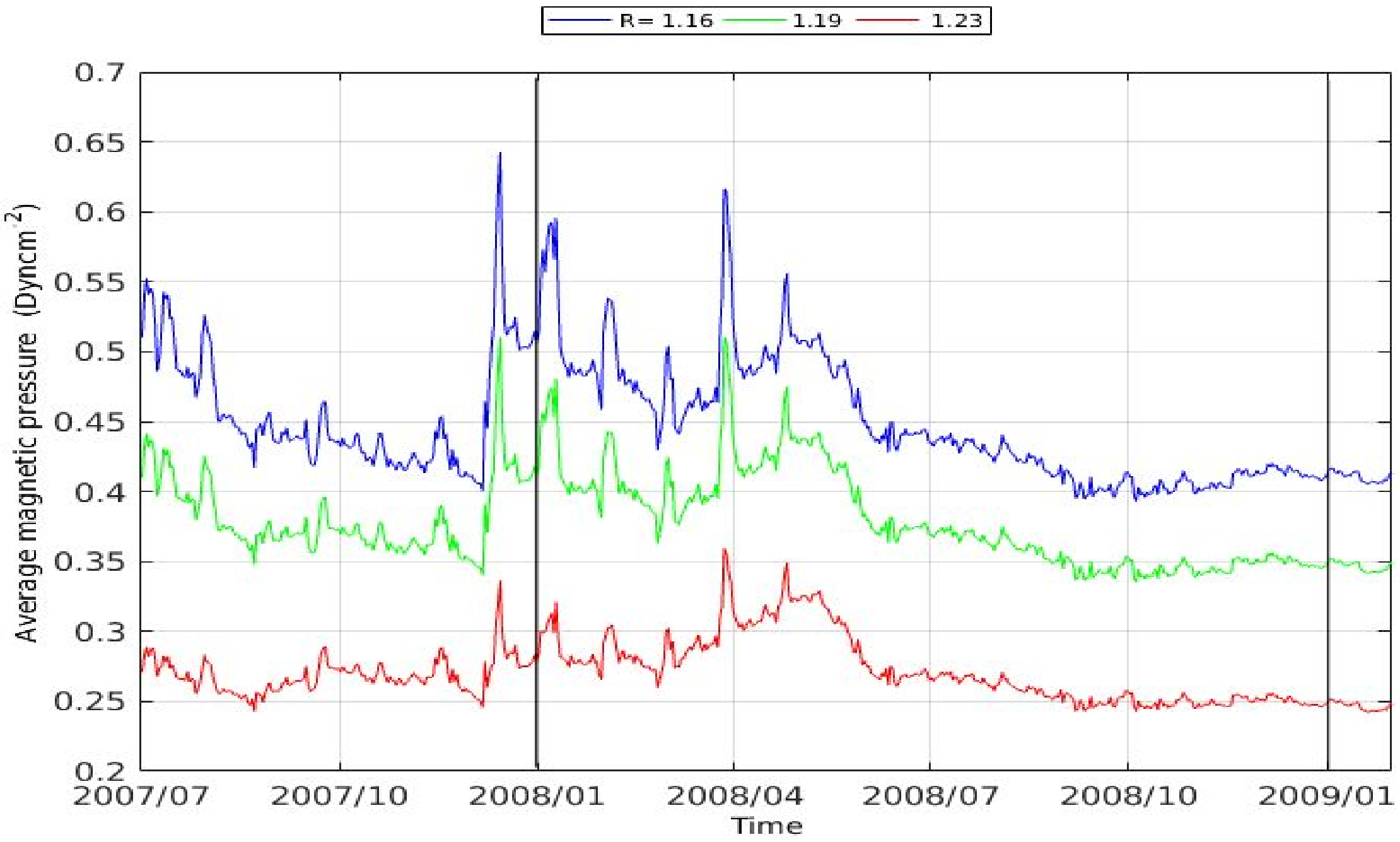

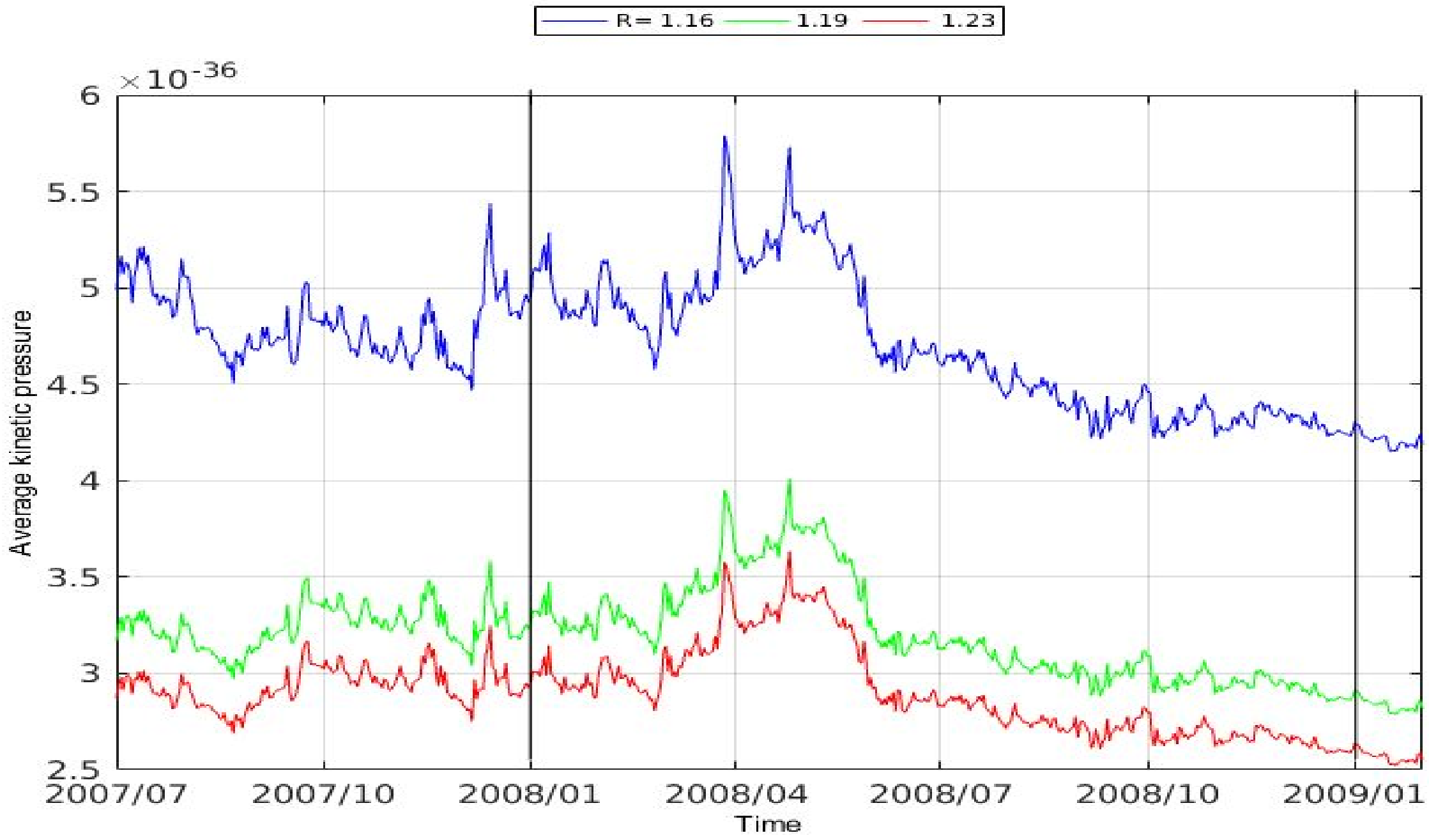

**Figure 5**. Variation in time, at each layer, of the average magnetic pressure (top panel) and average kinetic pressure (bottom panel) through different layers in the solar corona, vertical lines show time interval.

Figure 5 (bottom panel) is shown that the kinetic pressure presents an increase near to 2008/03 whereas after 2008/06 the kinetic pressure decreases. It is related to the average density variations in the same interval. In summary, the description from the CODET model shows a relationship between variations in kinetic pressure, magnetic pressure, electron density and magnetic energy during the period of interest. The increase of these quantities is correlated among themselves in the period from 2008/03 and 2008/06 and it shows small variations from 2008/07 to 2009/01.

In general, the temperature values obtained from these model descriptions are in accord with Osherovich et al. 1985. They describe a relation between magnetic, thermodynamic and dynamic structures in the solar corona during a sunspot minimum. In the same way, the CODET model assumes magnetic flux tubes in a stratified atmosphere similar to the approach in Osherovich 1984, but in our approach the magnetic field decrease with the height in the solar atmosphere. This behavior is due to the description of PFSS **B**-field lines.

In this Section, Figures 3 to 5 show the months-long slowly varying, yet under apparently stable conditions, rendered by the observed corona in a variety of light frequencies, e.g. those from the $Fe^{9,\ 12,\ 14}$ ions, providing i) statistical information on the corona temperature through their ionization temperature, as well as suggesting ii) a global state of equilibrium, constituting direct inference of two of several attributes of the quiescent Sun corona near solar minimum. These observations, as explained in this section, constrain the choice of the parameters in the successful model of the global values of density and temperature of the electrons in the 'low' corona region. Section 2 further introduced an 'indirect,' central quantity to the modeling. This is the intensity of the magnetic field in the Sun's corona, determined by the modeling of the magnetic field in the photosphere and inferred through the remote observation of the splitting of the excitation lines of He/O atoms, among other atomic element lines, see e.g. Zirin, 1966. There is also an interpretation of the manifestation of the magnetic field that is used which assumes it to be the one of an ensemble of flux tubes with an orientation dominated by being horizontal to the solar 'sphere.'

## 3. Complementary observations of the Quiescent Solar Corona

The optical thin nature of the quiescent Sun corona, see e.g. on the detection of the bending of a far start light, eclipsed by the Sun, see Dyson, Eddington, and Davidson, 1920, makes it difficult to identify the displacement of features, particularly when the density is close to homogeneous, i.e. differs locally in the region of consideration by less than a factor two or even a little greater than that. What follows is a checking of key observables which -although mostly located outside of the region of interest- guide us toward some of the possible conditions concerning matter and magnetic field in the overall solar corona altitude range studied here. In our case this is a volume contained in the spherical region delimited by $1.1 \leq r/R_\odot \leq 1.3$. Our goal is constraining the interpretation in Section 4 of the constitutive nature of the Sun's corona, which enables the CODET model introduced in Section 2 -of global hydromagnetic equilibrium- to be quantitatively successful in its predictions.

For the identification of the global properties of interest we take advantage of the observation remotely using: a) the coronagraph instruments for this and earlier solar minimum from SOHO and STEREO, b) for the non-thermal light that is measured by the EIT/SOHO instrument (Delaboudinière, et al 1995), and c) some of its large solar wind manifestations sampled in-situ from the Lagrange point L1 between Sun and Earth, with the monitoring of solar wind parameters plasma-protons (Ogilvie et al, 1995), and magnetic field (Lepping et al 1995), in our case using the Wind SC. We also use Earth's observations of the solar corona during the total eclipse.

### 3.1 About the magnetic field organization

In the quiescent sun corona, from 1.1 to 1.3 $R_{\odot}$ almost 100% of the source of the radiance, see Appendix 1, seems to be emission from a hierarchical arrangement of matter, see imprint of features in Figure 6 (a) and (b). These features, illustrated in Figures 6 (a) and (b), see e.g. Michels, 1998, appear to constitute an ensemble of magnetic flux-tubes oriented mainly parallel to the Sun surface, i.e. photosphere, and also see detailed studies of the Sun corona during the occurrence of total eclipse of the Sun in Voulgaris et al, 2010, Habbal et al, 2010, Daw et al, 2010. This is then consistent with one of the key assumptions by the CODET model, stated at the end of Section 2.

**(a)** **(b)**

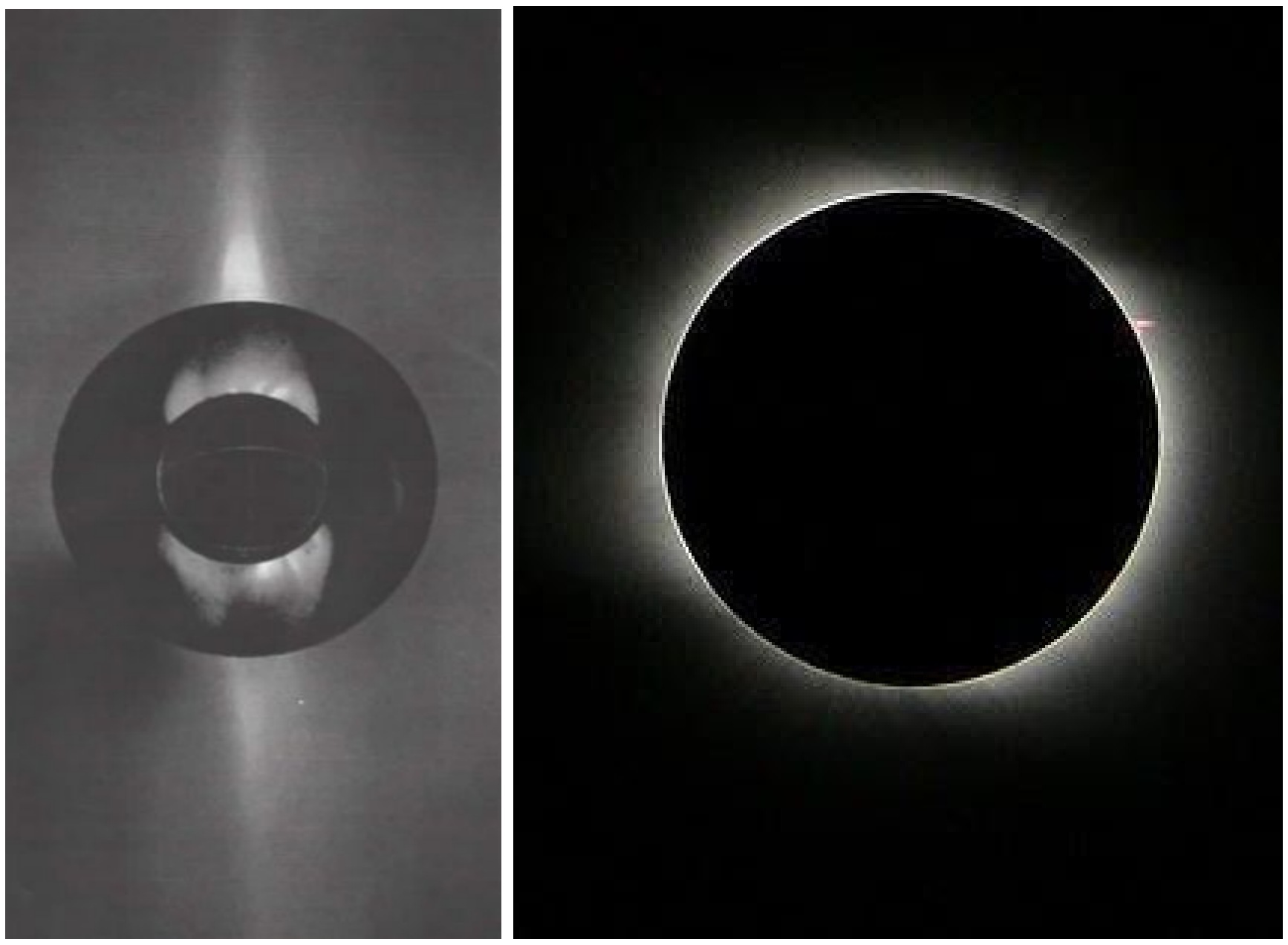

**Figure 6.** Magnetic field organization in the Solar corona. (**a**) Near solar minimum from Aug 1996, the solar plasma sheet self organization from its beginnings in the low solar corona, starting near 1.1 solar radius and extending up to 4 solar radii (Michels, D.J et al. 1998). (**b)** View of the low sun corona during the total solar eclipse of Aug 1, 2008, region of the corona discussed. View obtained at Novosibirsk, Russia.

Although difficult, it is possible to observe same as all pictures of the Sun presented, Figure 6 shows cumulated light along the line of sight. In the case of Figure 8a the observations are captured with the SOHO C1 and C2 near the 1996 solar minimum, see Michels, 1996, for details on the SOHO coronograph observations. Although difficult, it is possible to observe in the Sun corona, the temporal following of some fraction of arc-sec markers because of their luminosity in a relatively quick succession of coronograph pictures (e.g. every 12 min in coronal streamers, see e.g. movie of C2/LASCO month of January 1997[4] and observe changes in small plasma features at about 2 to 3 solar radii high in the corona, see e.g. Sheeley et al, 1997, Dere, Howard, Brueckner, 2000. This careful process offers some of the better delineated examples, i.e. markers suggesting their slow evolution in streamers toward higher and higher altitude becoming at some point the region that constitute a plasma-sheet, a characteristic feature of the nature of the slow to very-slow solar wind, see e.g. Vasquez et al 2017. A contemporary work by Viall and Vourlidas, 2015 says that the start of the SW, related to the quiescent corona in our time interval of interest takes place between 2.0 and 2.5 $R_\odot$. This SW tends then to possess specific properties when reaching a distance like the location of the orbit of Mercury. A SW that was in-situ observed by the two SC of the Helio mission, see e.g. Sanchez-Diaz et al, 2016, and which is understood to contain, for most of the time a current sheet in solar minimum that is like a disk surrounding the Sun close to its equatorial region. In this way Figures 6 suggest for larger distances from the Sun, e.g. beyond the one shown in Figure 6 (a), and which has been described up to 40 $R_\odot$, a set of contiguous thin equatorial streamers forming a streamer disk, in observations combining the SOHO coronographs C2 and C3 at solar minimum, see e.g. Michels, 1998.

In Figures 6 (a) we notice a certain asymmetry in the disk constituting the plasma sheet during the period of the observations presented. A likely cause of that asymmetry is considered below in the sub-section 3.3.

In consequence it is important to consider that at solar minimum in the Sun corona there appear the following characteristics: a) streamers, b) quiescence, c) apparent equilibrium, d) reduced presence of weak transitory (e.g. flares, filament disappearances) associated or not with sudden ejection of magnetic field and matter that tend to displace at slow speed when compared with the historical mean value of the solar wind convection speed (~ 400 $kms^{-1}$) away from its source, see e.g. Priest, 2014. A source that is possibly located at about between *2* to $8\ R_0$, depending on the solar corona conditions, see e.g. Schwenn and Marsch, 1990, and ≈ $2\ R_\odot$ following SECHI/STEREO observations at solar minimum of interest by Viall and Vourlidas, 2015. Notice, SC Solar Probe perihelia is planned to be as close as $8\ R_\odot$ in/near 2024.

[4] https://sohowww.nasa.gov/gallery/Movies/series.html

### 3.2 Signatures of the EUV light

During the time interval known as solar minimum, the time when the quiescent sun corona appears to dominate we quote '… the extreme ultraviolet EUV light, monitored to high resolution by the EIT instrument in SOHO shows the reduction in features other than the super-granular well-known arrangement of the Sun's photosphere which contains the rising energy toward the chromosphere and beyond, possibly represented in the form of spicules, rich in the magnetic field and neutral as well as plasma matter from below the photosphere region,' see De Pontieu et al, 2009, 2011. Gomez, Bejarano, Mininni, 2014 present an interesting view of magnetized, fully ionized plasma conditions that we assume may prevail in spicules type II 'delivering magnetized-matter at/above the chromosphere/corona transition region (TR) heights.

In this subsection, we check some features of the solar corona and their evolution during the solar minimum. In this case, we used the EIT/SOHO images in the EUV band, The usual corrections were applied under solarsoft, i.e. EIT images in 19.5 nm, 17.1 nm and, 30.4 nm are selected on April 01 (2008), about 14 months away from solar minimum (Figure 7). In this period, is possible to see three ARs (NOAA 10987, 10988 and 10989). The solar corona evolution is illustrated next with about a year to the solar minimum of selected pictures, and starting on Jun. 01 (2008). Here we show four images in EIT lines (in Figure 8). It is possible to observe that besides a supra-granular network, still, a few features like small AR appear present. The presence of these structures seem somehow reduced at about 6 months from the solar minimum when a whole set of 4 EIT line images are shown in Figure 9 (Nov. 01, 2008). Finally, Figure 10 shows in four images (Jan. 01 (2009) the closest, 4 months to solar minimum conditions in the low corona. Figures 7, 8, 9, and 10, in summary, show a march in time of the solar corona toward solar minimum, in which few features stand out, besides the fact that in Jan 2009 the presence of CHs of which the most prominent are near the poles.

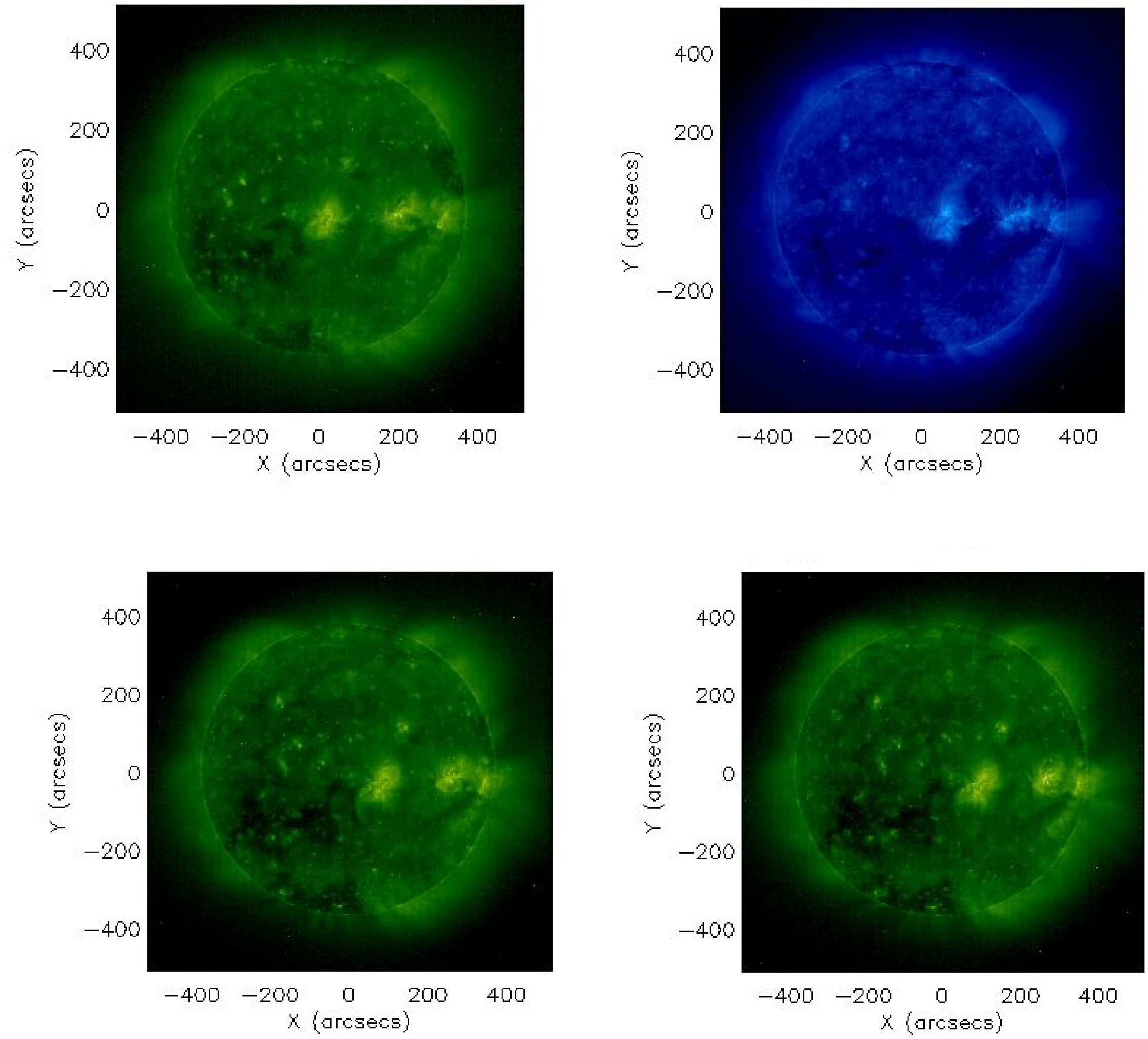

**Figure 7**. EUV Images from EIT/SOHO in April. 01 (2008). Upper left panel 19.5 nm image at 01:06 UT. Upper right panel 17.1 nm image at 13:00:13 UT. Lower left panel 19.5 nm image at 19:25:05 UT. Lower right panel 19.5 nm image at 23:48:09 UT.

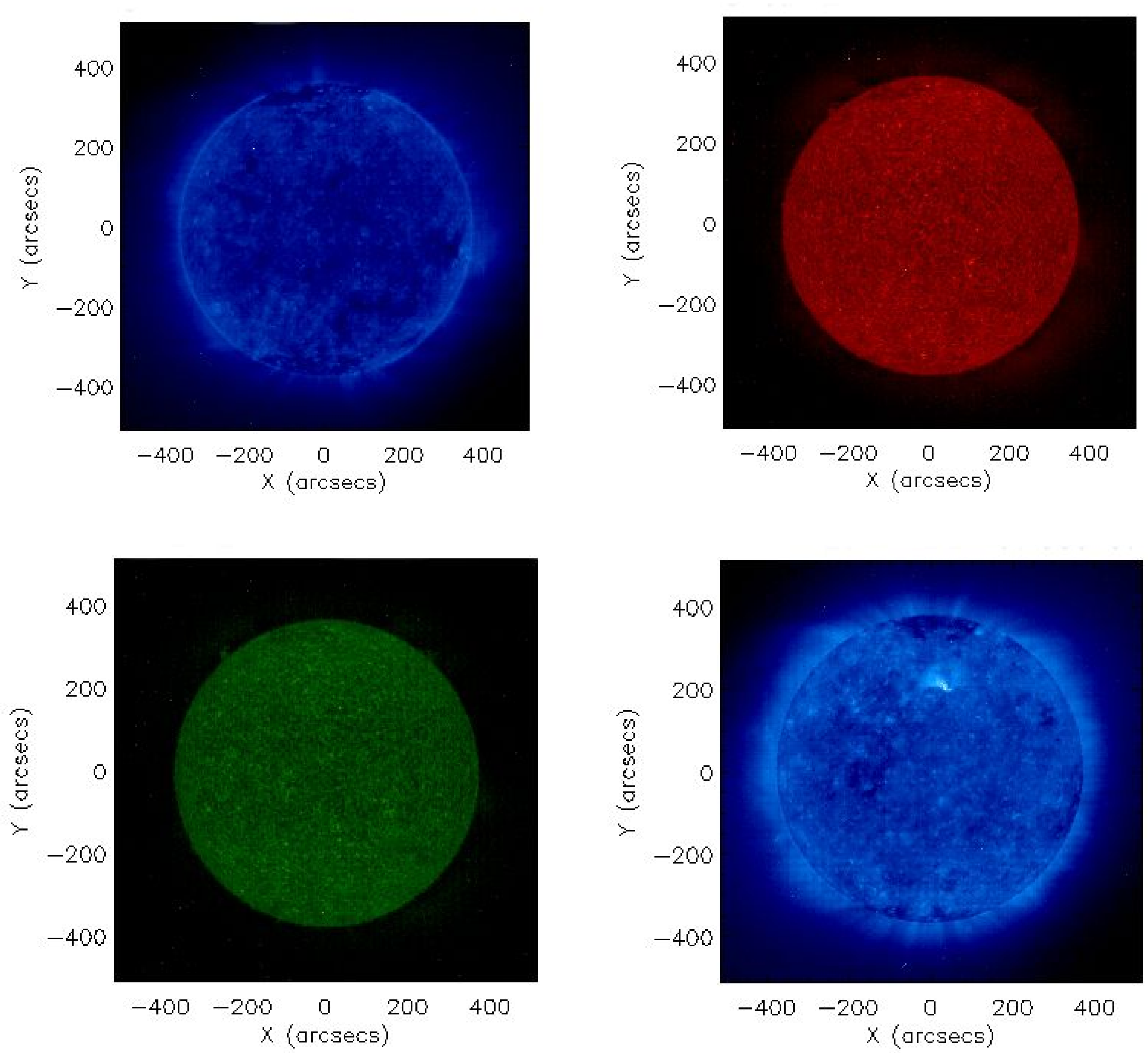

**Figure 8**. EUV images from EIT/SOHO on Jun 01 (2008). Upper left panel 17.1 nm image at 01:01:13 UT. Upper right panel 30.4 nm image at 07:19:34 UT. Lower panel 19.5 nm image at 23:45:07 UT.

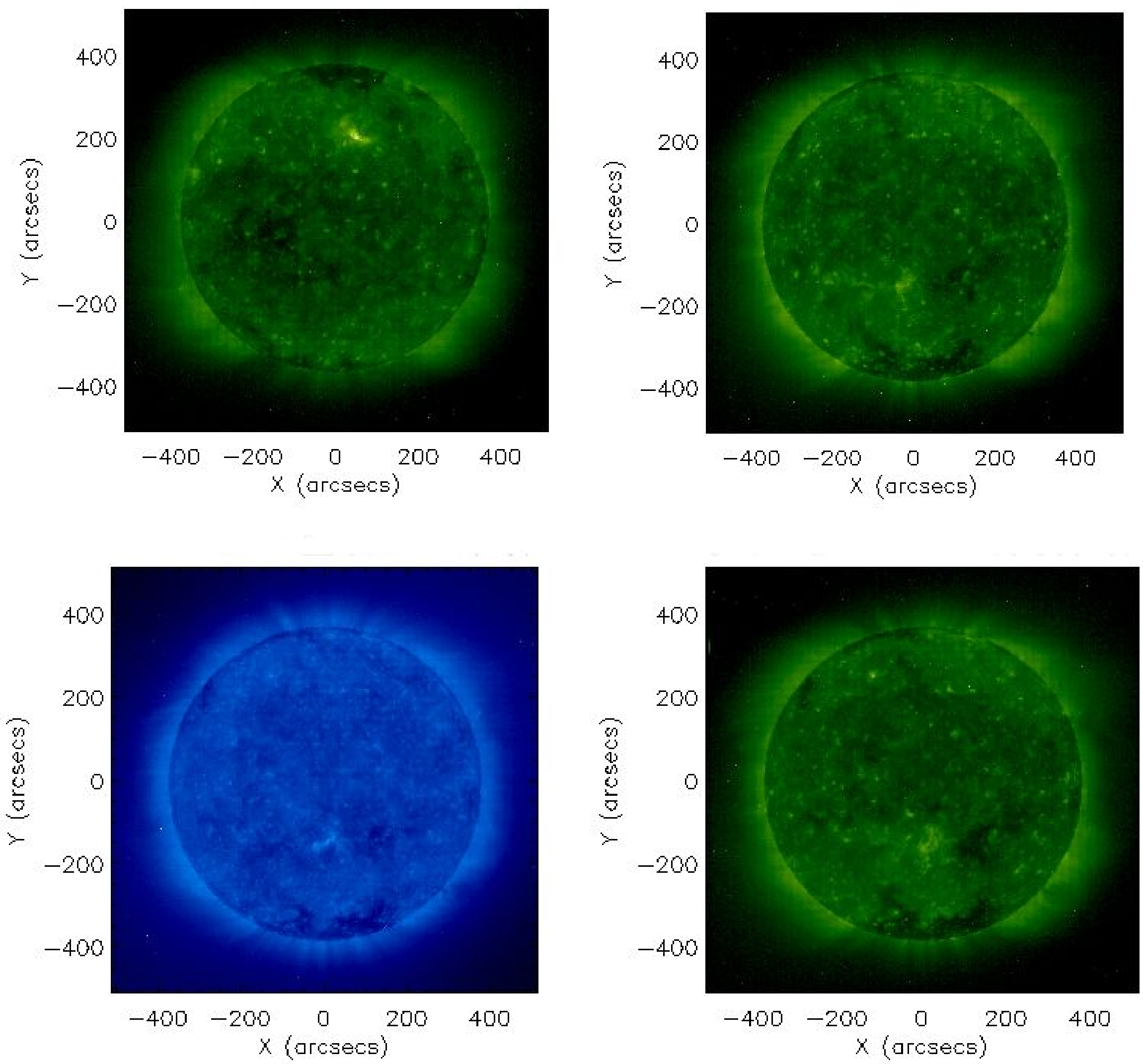

**Figure 9**. EUV images from EIT/SOHO on Nov 01 (2008). Left upper panel 19.5 nm image at 00:00:11 UT. Upper right panel 19.5 nm image at 12:24:10 UT. Lower left panel 17.1 nm image at 18:00:13 UT. Lower right panel 19.5 nm image at 23:36:09 UT

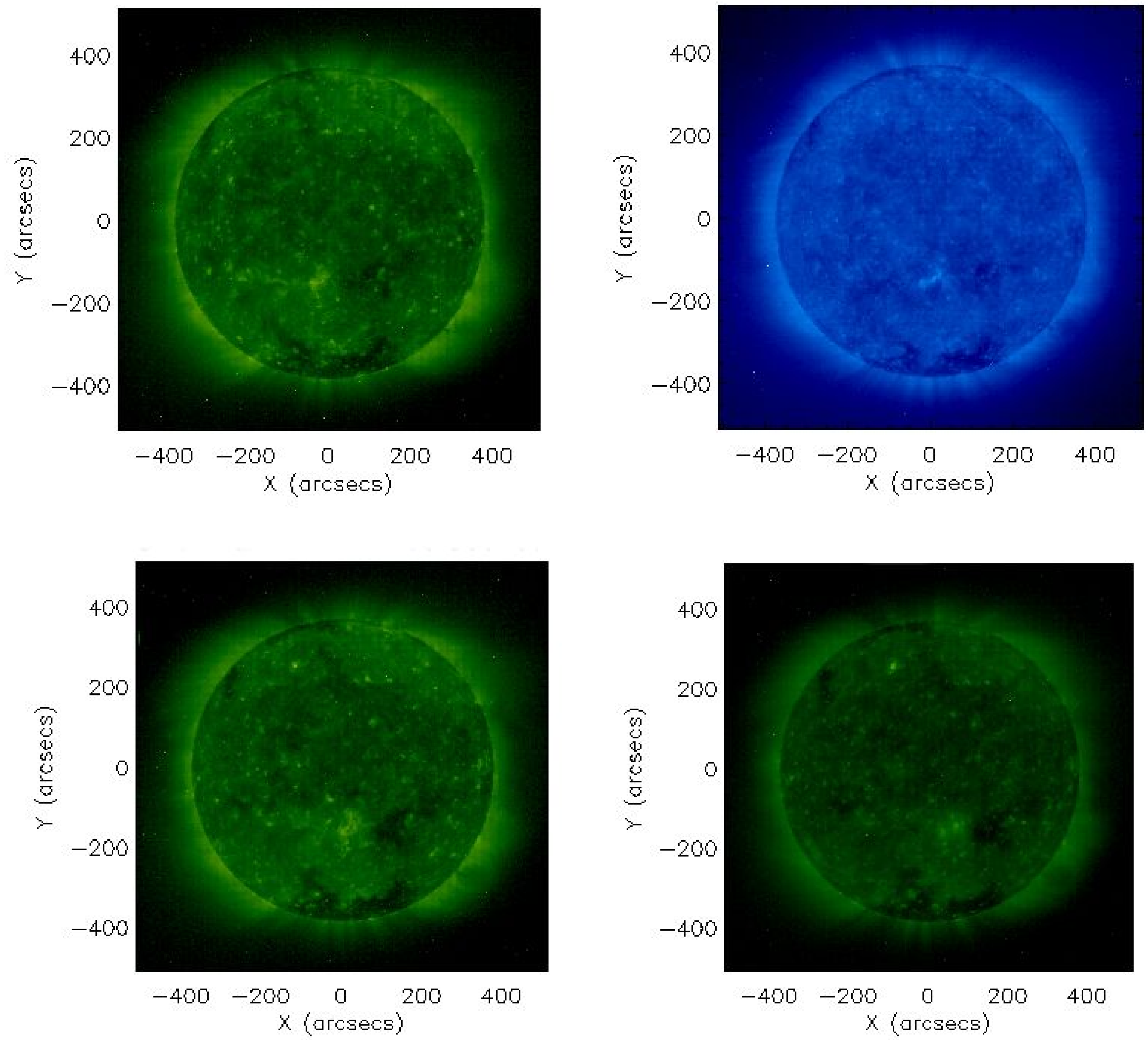

**Figure 10**. EUV images from EIT/SOHO on Jan 01 (2009). Left upper panel: 19.5 nm image at 00:00:08 UT. Upper right panel: 17.1 nm image at 07:0012 UT. Lower left panel: 195 nm image at 12:48:09 UT. Lower right panel: 19.5 nm image at 23:36:09 UT.

The minimum between cycles 23 and 24 is particularly interesting. Its temporal extent, it is not seen before during the space-era of sophisticate exploration of the Sun and its environment remote and in-situ for some of its solar wind larger manifestations. This solar minimum exhibits unprecedented properties shown in figures in section 2.6. At solar minimum times, the GOES under-counting effect for weak flares becomes less significant. But it appears to be the case that CMEs continued to occur, although large flares basically ceased in 2008 (only one M-class and 8 C-class flares, see e.g. records stored by NOAA at its World Wide Web location[5]).

---

[5] ftp://ftp.swpc.noaa.gov/pub/warehouse/2008/2008_plots/2008_xray.tar.gz

It is pointed out by Webb et al. 1998; Robbrecht, Patsourakos, Vourlidas, 2009; Hudson & Li 2010, that the weaker coronal magnetic field at minimum times is easier to disrupt, in consequence, the CMEs can occur in major frequency. This is an assumption that appears to be corroborated by changes in the frequency of observation of solar transients in the SW at the Earth's Lagrange point L1, see e.g., Lepping et al, 2011.

### 3.3 The 1 AU (in-situ) Manifestations and its Remote Solar Corona $N_e$, $T_e$

SW high-speed stream(s) observed at heliospheric ecliptic region at 1 AU, in the Figures 11 and 12 show the imprints of solar corona equatorial holes likely present at all times in the solar minimum period low-latitude, although strongly reduced or absent near the actual minima (June 1996 in a previous solar minimum, and April 2009 in the most recent one, see e.g. Steinhilber, 2010). Indeed at Carrington rotations containing solar minimum co-rotating high-speed solar wind cease or their speeds are highly diminished as the Carrington figures[6] for them show in Appendix 2. In the interval of our study, we notice that sporadic sunspots seemed to produce more complex, and long-lived active regions.

Figure 11 illustrates for April – May 2008 a more complex, both in average and mean faster SW, than the later observed simpler overall SW in Figure 12 from July – August 2008. It is worth noticing that the circumstantial greater complexity in the SW matches the more complex changes in density $N_e$ of Figure 1 for the solar minimum and specific to this study Figure 3 for April – May rather than after June 2008 when there are simpler smooth variations in $N_e$ that reach in about nine more months the lowest values –solar minimum– on or about April 2009. Conditions starting about June 2008 are the ones that match the simpler observed conditions of the SW of Figure 12. Be warned that global remote sensing of the quiescent Corona here is compared to a single in-situ measurement at the Earth's L1 location in the SW.

[6] https://cdaweb.gsfc.nasa.gov/cgi-bin/gif_walk?plot_type=wind_27_day_kp_plots

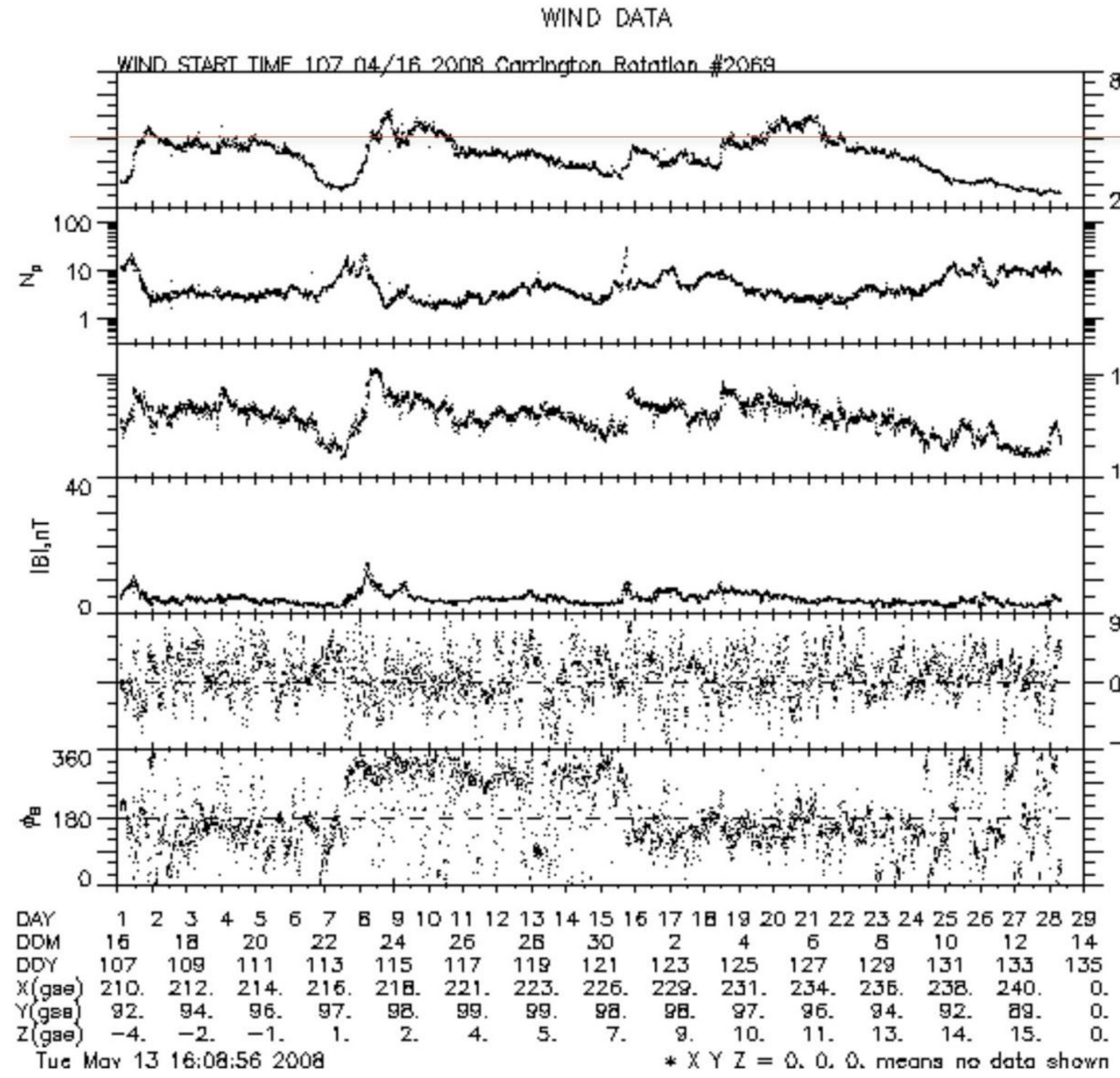

**Figure 13**. As a function of time (Carrington rotation number 2069, starting time on May 16, 2008) shown are the solar wind 'key' parameters collected at 1 AU by the Wind SC; (1) |**V**|, convection velocity of the magnetized solar wind plasma, primarily away from the Sun, (2) $N_p$, proton density, (3) $V_{th}$, thermal proton velocity, corresponding to a $T_p = V_{th}$ sqrt(k/2$m_p$), (4) |**B**|, the magnitude of the convected, in the plasma frozen, magnetic field, (5) $\vartheta_B$, the angular direction of **B** respect to the ecliptic plane, (6) $\phi_B$, the angular azimuth direction of **B** in the ecliptic plane. The bottom-top line counts the number of days included, 2nd bottom line reports the day of the month (DOM), 3rd row gives the day of the year (DOY), 4th row indicates in X-axis along with Sun – Earth line the SC Wind location from the center of the Earth in Earth's radius, completing the location of the SC Wind 5th and 6th lines give in GSE coordinate system the location in the Y along azimuth positive in the direction of the Earth motion around the Sun, and Z positive for North from the ecliptic, completing in this way a right-hand coordinate system[7].

[7] https://cdaweb.gsfc.nasa.gov/cgi-bin/gif_walk
Plot type wind 27 days survey, specific date 2008107

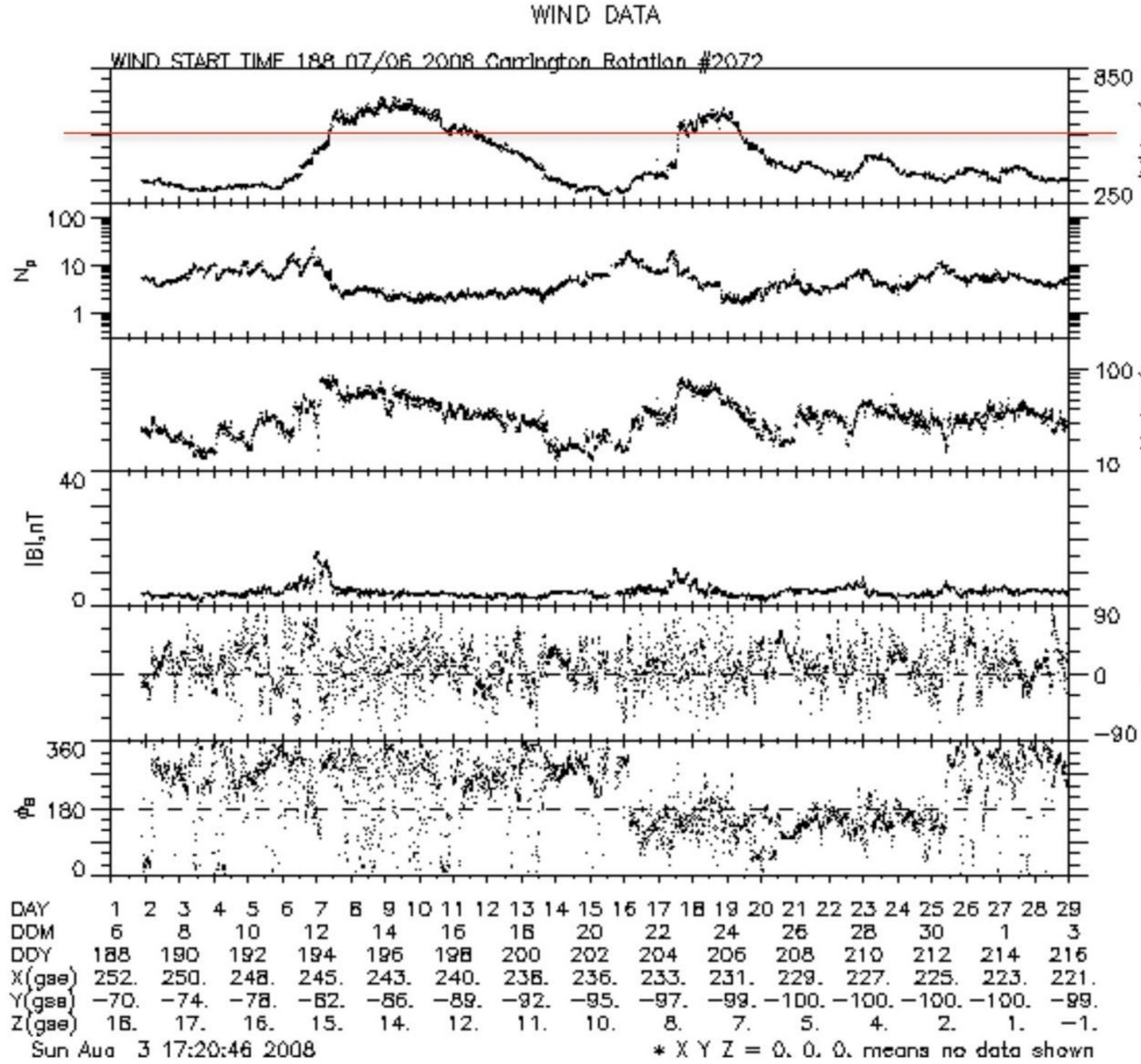

**Figure 14**. Same as Figure 13 but for the Carrington rotation number 2072, starting time on July 6, 2008[8].

When comparing in-situ observations with conditions in the quiescent solar corona it is relevant to keep in mind that the location of the disk at the solar minimum region with the bulk of plasma emanating from the quiescent Sun corona possibly concentrates in a narrow region of space beyond several solar radii centered at the Sun equatorial plane. The Sun equatorial and the ecliptic planetary planes intersect solely in the first week of June and December. The observed at the solar minimum in 1996 and 2009 SW speeds at 1 AU in the ecliptic plane are consistent with this understanding of a narrow disk of plasma with this slow SW at the solar equatorial plane. For more definitive connections between SW observations and the quiescent Sun corona it will be useful to have a thorough study beyond the scope of the present one with simultaneous in time observation at other locations of the SW as they are available from STEREO,

[8] https://cdaweb.gsfc.nasa.gov/cgi-bin/gif_walk
Plot type wind 27 days survey, specific date 2008189

Messenger, and Luna Observer and possible other missions, Solar Probe recently launch will be of great help in this regard.

Solar irradiance in the interval of time between 2008-2009 appears to indicate that the flux emergence reaches a lower level; while the existing magnetic flux flow evolved causing changes in the large-scale magnetic field that seem to diminish the presence of the equatorial CHs as solar minimum approached, see e.g. Richardson and Cane, 2012. In conclusion we understand that the in-situ high-speed streams on 2008 confirm that there was in the Sun corona altitude range of this study a presence of *'unipolar'* magnetic regions at mid to low latitudes related to the CHs (Schrijver et al. 2011; de Toma et al. 2010). These CHs appear to directly relate to the lowest EUV radiation values present in some regions of the low corona, CHs regions that appear to decrease from Jan 2008 to Jan 2009 in Figure-sets 7 to 10.

That decrease in CHs area suggests (the source of the fast solar wind) diminishing as time pass, see Appendix 2. And that decrease expresses in the reduction of the high speed stream regions in Carrington rotation as shown in Figures 11 and 12 as well as the added intervals in Figures A2.a and A2.b. This result interpretation enables our understanding that irradiance collected from 1 AU since the start of year 2008 is marginally sensitive to CHs. I.e. the *irradiance is dominated by the one generated from the quiescent* corona with a relative higher radiance values than in CHs regions. This further implies that in the altitudes of interest irradiance is sensitive mainly to the *parallel to the Sun surface field lines.* Figure 6.b shows remote direct measurement at the total Solar Eclipse, August 2008, which as discussed are supportive of the CODET model prediction of the corona $T$ and density ($N_e$) profiles. These time dependence profiles of $N_e$ show negligible change after June 2008 (Figure 3), and temperature ($T_e$) profiles do not show more significant changes already since April 2008 (Figure 3). However, the average energy density (Figure 4) related to the variations of the magnetic field show an increase in April 2008, in agreement with the behavior in the solar atmosphere, that some ARs appeared in this period (Figure 7). In the same way the average magnetic and kinetic pressure shows an increase in April 2008 (Figure 5).

Notice that in our one year time interval, the final period June 2008 to January 2009 shows in Figure 3 to 5 smaller than before variations in density, $N_e$, average electron energy density, average magnetic and kinetic electron pressure, and a constant temperature, $T_e$ related to features of the quiet corona which are central to the interpretations in the simple thermodynamics models made in Section 4.

### 3.4 Topological thermal insulation and other Estimates of Temperature in the Quiescent Sun Corona

On the insulation between the quiescent corona and other features with a different temperature shows in Nikolsky et al., 1971, near the time of a solar maximum, a clear example describing conditions in the region of the Sun corona, between a quiescent corona with T > $10^6$ K and a prominence having a much larger mass, and temperature two orders of magnitude smaller ($10^{11}$

e/cm$^3$, and $10^4$ K). Another interesting aspect of the study is the observation of non-thermal contributions to the width of Sun corona non-thermal lines. The obtained velocity, when split between the line of sight and transverse to that at the base of the corona was (10, 25, 25) kms$^{-1}$. During the eclipse was estimated a great degree of inhomogeneity in the e-density in the 'quiescent corona' suggesting a possible turbulent activity. Here we speculate that it could also be possible that there was a contribution to the inhomogeneity from some unresolved streamers for this thoroughly described eclipse that corresponded to the active part of the solar cycle, March 7, 1971.

Under the solar minimum conditions of August 1, 2008 total solar eclipse, we found the work by Habbal et al, 2010 in which a thorough discussion is devoted to an interval in altitude in the corona that contains this study's range. Their work distinguished between two essentially different regions, the one near the poles where with help of the Fe X and XI a temperature close to $10^6$ K is identified. This result appears to be in agreement with the high-speed SW, out of the ecliptic measured Fe frozen charge states in Gloeckler and Geiss, 2007. For the quiescent corona the temperature identified in Habbal et al, 2010 is consistent with the CODET model illustrated in Figure 3.

### 3.4.1 The low Solar Corona Temperature from thermal equipartition of Iron ionization states

Defined from emissions in lines of iron the analysis of the observations of the intensity of the Fe X and XIV charge states during the total eclipse of August 1, 2008, observed in Novosibirsk the detailed analysis in Voulgaris, et al, 2010 concludes that the observed radiances of Fe XIV set an upper limit to the quiescent Sun corona temperature of $1.9\text{x}\ 10^6\ K$. The strong radiance of the Fe X line set a lower boundary to the corona temperature of $1.2\text{x}10^6$ K. This is a range that contains the value that the CODET model gives to the electron temperature in the region of the study during the year 2008. For other work on the solar minimum in which a more complete description of Fe charge states exists, see e.g., Habbal et al, 2010, Daw et al, 2010.

In what we consider is the bulk of the irradiance in our study's temperature during the total eclipse of August 1, 2008 Habbal et al, 2010 find that the thermal temperature of the quiescent corona we have a value constrained between 1.6 and $1.9\text{x}\ 10^6$ K. In particular, during the eclipse of August 1, 2008 Habbal et al, 2008 divide the non polar coronal hole regions into regions where there are streamers and no streamers identified at the base of the corona. The regions classified as non-streamers appear to be harder to observe having a weaker signal in the Fe charge states. Still, in both cases, an estimate for the electrons temperature suggest the observation of mainly Fe XIII and XIV non thermal emission indicating a temperature similar to the one observed in the streamers at altitude $\leq 1.3\ R_\odot$.

### 3.4.2 The temperature of the low Solar Corona from the width of the iron non thermal lines

Using width of optical lines in iron isotopes non-thermal light emissions in coronograph C1 in SOHO, Mierla et al 2008, present a study between August 1 and October 31, 1996, an interval containing the minimum between solar cycle 22 – 23. The study shows that in the Sun corona between 1.10 and 1.80 $R_{\odot}$ the Fe X 636.6 nm signal appears to be wider in most cases than the Fe XIV 530.3 nm emission. The consequence of this result suggests that non-thermal effects contribute to the increase in the width of this irradiation line Fe X 637.6 nm, which belongs to a charge state of iron that occurs when its temperature is $10^6$ K. Fe XIV is a charge state that occurs when the temperature of the medium is 2 x$10^6$ K.

Further the width of the Fe X and XIV show consistently a small gradient in the height range of our interest, i.e. the temperature predicted appears close to constant in the 1.10 and 1.30 $R_{\odot}$ height range, as it is shown in Mierla, et al, 2008 Figure 8 right panel. This appears consistent with the CODET prediction for the solar minima of the same temperature in the range discussed from 1.16 to 1.23 $R_{\odot}$ .

Finally, if we assume an equal-partition for the Fe charge states, and considering the two states discussed in this work we obtain–from Figure 1 top two panels in Mierla et al, 2008 – a value for the temperature given by the relationship:

$$T= \frac{(A*2x10^6K+B*10^6K)}{A+B}$$

where A = radiance(Fe XIV– continuum, 530.3 nm) = (174 – 163) radiance/s,

and B = radiance(Fe X– continuum, 637.6 nm) = (98.5 – 95.6) radiance/s.

and the resulting temperature is then T = 1.75 x$10^6$ K for the region 1.13 – 1.25 $R_{\odot}$ .

Although this temperature estimate makes strong assumptions that are at best reasonable under extremely simple conditions of overall homogeneity, it is interesting that it is in quantitative agreement with the CODET prediction from Figures 3 for the model presented in Section 2.

### 3.5 Interpreting the remote and at 1 AU in-situ manifestations

We start by citing a large portion of the abstract in Michels, 1998 which fundamentally contains a good portion of the interpretations, that we reach from the analysis of observations in sub-sections 3.1 to 3.4, i.e. we closely agree with Michels in key elements with the views reached in the observation that:

"… the global corona at the minimum phase of the Sun *current* activity cycle show an organization that is striking, both in its simplicity, and its persistence. Time series of images from the four telescopes that comprise the LASCO and EIT images on the SOHO satellite, show the classical solar minimum dipole configuration of the outer corona. But in addition careful analysis makes clear that (1) the low lying closed-loop multi-polar magnetic structures characteristic of the inner corona at low to mid-latitudes, are but the lowest tier of a hierarchy of loops that stretch across ever greater distances as we observe at higher and higher altitudes in the solar atmosphere, (2) these nests of loops are not static, but stretch continually outward, and in the process feed plasma into the higher structures and hence to the solar wind, and (3) that, at this phase of the activity cycle ..."

We conclude with a listing of the physical interpretation of the remote observational findings of the sub - Sections 3.1 – 3.4, as well as at 1 AU in-situ manifestations, which in Section 4 are taken into account in our model interpretations:

1. The quiescent solar corona in the region of interest appears to be constituted by an *idealized* ensemble of magnetic tubes contained in regions, Section 3.1, which are consistent with the assumption of having at their micro-scale level a cylindrical shape, being in an orientation parallel to the Sun's photosphere. Here, we assume them to be regions homogeneous in the limit of the observational resolution capability, i.e. with a cross section of about $\pi l^2$ and length ($4l$), *with* $l \approx 25\ km$, from a rough. Although an estimate at the Sun's corona near chromosphere/TR from Berdichevsky and Schefers, 2015. In section 4 we will use a value $l = 20\ km$.

2. It is simple to extrapolate, see Appendix 3 on the ratio of SW flowing in the streamers away and without return to the solar corona mass at altitude 1.23 $R_\odot$, that the SW contains just a fraction of no more than 1/16 of all mass in the corona region of interest, *1.1 < r < 1.3*$R_\odot$, which is consistent with that region at solar minimum as being the source of the SW (Figures 11 and 12) from the equatorial solar streamers (Figures 6) which exist in the heliosphere plasma-belt enclosing the Parker current sheet. This result lets us assume that the quiescent solar corona considered contains mainly 'matter trapped in closed magnetic field tubes.' This is again consistent with the suggested interpretation of the properties of the matter and magnetic field expressed in the point before (1).

3. That, as discussed/illustrated in Sections 3.2 and 3.4; There is for the times studied a very small presence of filament/prominence eruption. We considered there to be no recording of any observation in the Sun-Earth line of sight of any such flux-rope structure from Dec 2007 to Dec 2008, the time of the study. A subgroup of all ejecta is discussed in Lepping et al, 2011, Also their presence, quite common for all of 2009, did not seem to affect in any substantial way the prediction made for the nature of the electrons density and temperature for more than 2-years of quiescent Sun corona at the solar minimum of 2009. A more extended interval of identification of solar ejecta of the magnetic field dominated kind containing this solar minimum is in Wu and Lepping, 2015, see also Appendix 4.

4. That although for most of the quiescent intervals there is a presence of some Sun's Coronal Hole (CH) region(s) with a substantial impact at 1 AU, see Section 3.3 on the generation of a fast speed solar wind, which it is possible to document from in-situ observation of the SW at L1. It seems we do not need to address the fundamentally different nature of their thermodynamic properties in this study because the source region is likely of lesser relevance to the global irradiance from *1.1* to *1.3* $R_{\odot}$,. (Notice how no signatures of a quantitative influence appear in the CODET model predictions for $N_e$ and $T_e$ across a few Carrington rotations when SW observations show the transition from intervals with high speed SW to those with slower speed SW while reaching the solar minimum in April 2009, thereby implying an *absence* of high speed SW is also an *absence* of equatorial CHs precisely near the solar minimum in June 2009.)

5. That the model electron density and temperature found in Section 2 show reasonable agreement with two different direct inferences on $T_e$ presented in Section 3.4.
6. It is concluded from all the considered observables discussed in Section 3 that there exists an active mechanism responsible for the insulation of the quiescent low Sun's corona of our interest, from the other in this section mentioned different structures at the same altitude and with different thermal and magnetic properties, i.e., i) equatorial Coronal Hole(s), ii) filament/prominence(s), iii) small ARs and/or ARs remnants.

The fact that thermal insulation occurs in the superior regions of the Sun is dramatically proven during the solar minimum when for extended quiet intervals during the sun activity there is a well-known massive temperature difference between the photosphere and low corona with dramatic variation in value in the TR, just passing the chromosphere, see e.g. Vernazza, Anrett, and Loeser, 1981.

The models (next section, Section 4) present interpretations of the constitutive nature of the magnetized matter that populates the corona between 1.1 and 1.3 $R_{\odot}$ and consistent with indicated observations in Sections 2 and 3.

## 4. Thermodynamic interpretation

### 4.1 The CODET model implication on the relation between $N_e$ and $T_e$

Sections 2 and 3, including Figures 1 to 5 show that the model presented in Section 2 represents essentially a valuable proxy capable of describing empirically the steady presence in the corona of high temperature and the quantitative estimate of the plasma density with a narrowly constrained uncertainty. Further Section 3 was devoted to making clear that in doing so the CODET model in Table 1 is capturing properties almost exclusively characteristic of the quiescent solar corona in a region in altitude that extends from $1.1 < r/R_{\odot} < 1.3$. Here, *r* is the distance from the center of mass of the Sun. It is assumed for all purposes to coincide with its geometric center $R_{\odot} = R = 7\text{x}10^5$ km*,* corresponding to the average location of a rather *thin* photosphere, about *1000 km* thickness, see e.g. in Vernazza, Avrett, and Loeser, 1981. Assuming then an ideal gas of electrons in the corona the identity connecting electrons number $N_e$ and $T_e$ to the **B**-field magnitude is identified as

$$N_e = B^2\beta/(16\pi\mu k_B T_e) \tag{8}$$

where

$$\beta = p/(\tfrac{1}{4\pi}B^2/2\mu) \tag{9}$$

It is assumed *i)* to be a constant, *ii)* a property of the quiescent Sun corona medium at each altitude layer, and *iii) ab initio* assumed an ordered stratification of the medium in altitude. Then

$$\beta/16\pi\mu K_B T_e = f(B) \tag{10.a}$$

and

$$(16\pi\mu K_B N_e)/\beta = g(B) \tag{10.b}$$

They are valid, i.e., simple relationships with $f(B) = Ct\ B^{2-\zeta}$ and $g(B) = Ct\ B^{2-\alpha}$. Notice that the empirically obtained values for $N_e$ and $T_e$ when using the vacuum permeability $\mu_0$ imply a plasma $\beta = 2.5 > 1$.

*A*ssumptions expressed in Equations 10 are explored in the model by using the measured photosphere **B**-field magnitude extrapolated to the location in the corona where electrons $T_e$ and $N_e$ are extracted producing estimates within an uncertainty of about 20 and 30% respectively (as it is identified in sections 3.4 to 3.5). From Equations 8 to 10 we arrive at the presented expressions

$$N_e = B^{\xi}/Ct_1 \text{ and } T_e = B^{\alpha}/Ct_2 \tag{11}$$

The step by step details are given in Rodríguez Gómez, 2017, Rodríguez Gómez et al, 2018, as well as in a more concise version in the Section 2, which includes the parameter values for this model listed in Table 1. The model gives an altitude ($r - R_\odot$) dependence for $T_e$, and $N_e$ when the assumptions presented in Equations 8 to 10 are used.

### 4.2 A possible interpretation of the model polytropic index *γ*

First we notice from Equations 11, that the relationship found in the empirical model of Rodríguez Gómez. (2017), connecting temperature and density of the electrons in the corona is

$$N_e \propto T_e^{\gamma-1} \tag{12}$$

where the thermodynamic parameter $\gamma$ for an ideal gas adiabatic process (electrons in this case) is, so far, simply a parameter that depends in the model value of $\zeta$ and $\alpha$ as shown in the Appendix 5. Then the relationship from Equation 12 appears valid, as illustrated in Figure 1, because as the model introduced here shows, the parameters in Table 1, adequately describe the electrons number ($N_e$) and temperature ($T_e$) for a substantial part of the Sun corona, extending from about 1.16 to 1.23 $R_\odot$.

It is relevant then to point out that Equation 12 identifies an adiabatic condition where an ideal gas is subject to a thermodynamic reversible process, while entropy ($S$) is preserved. However, the relationship needed to describe qualitatively the Solar corona observations, Rodríguez Gómez, 2017, requires $\gamma \simeq 1/2$, while the quiescent corona solar minimum CODET model, see *Table 1* values for $\alpha$ and $\zeta$, imposes a $\gamma \simeq 1/6$. Both values of the polytropic index are smaller than one. Therefore, Equation 12 describes an adiabatic process when the gas density decreases the gas heats up (i.e. $\gamma < 1$).

In addition notice that for the above obtained $\gamma$, following the derivation in *Section 2* in Berdichevsky and Schefers, 2015 we found that the anomalous polytropic index $\gamma = 1/2$ corresponds to a parameter $\eta = 7/4$ *that is* the coupling factor between the competing e-gas pressure and magnetization works. In the present case $\gamma = 1/6$ *corresponds to a coupling parameter* $5/4$ *of the magnetization work to the e-gas work.* (See derivation in Appendix 5 and in the listed values of Table 2).

Next, we proceed to discuss how our interpretation of a *magneto – matter – corona structure* is consistent with the empirical relation identified in Table 1 for the description of the quiescent 'low' sun corona considering that the relationship from Equation (12), i.e. $N_e \propto T_e^{\gamma-1}$ appears valid.

It will require a theoretical interpretation, that we develop, to explain a diamagnetic condition consistent with the above assumed polytropic $\gamma < 1$ in the Sun Corona. Naively we would also expect a low plasma *β* to be the case. However, it is not established in our region of interest that the plasma *β* is low. This happens although this interval of time is of the quiescent nature of the low Sun's corona region at an extended solar minimum. Consider that, while |**B**| changes are at most a factor among photosphere and the low Sun's corona, particles number for this likely fully ionized medium goes from N ~ $10^{15}$ at $R_\odot$ to ~ $10^8$ at r ~ 1.1 $R_\odot$ and even less at solar minimum as Tables 1 and 2 show, see also Figure 4. However, at the same time that $N_e$ decreases the $T_e$ increases in an comparable order of magnitude (see low corona $T_e$ in Figure 3).

By considering the conditions of the medium to be in thermal equilibrium in the region, we proceed from now on to drop the sub index 'e' from the temperature 'T.'

As alluded to in Section 4.1, the study is constrained to remote sensing inferences only. Such a situation introduces higher uncertainties both in observation and in interpretation. This is so because the properties of the medium associated with $\gamma$, substantially smaller than 1, cannot be explored in-situ. This is due to the fact that the materials, we use for observations are heat conductors and do not resist the destructive effect of the observed temperature/radiation so near to the Sun's photosphere/ chromosphere.

**Table 2.** Magneto-matter constant values resulting of the thermodynamic interpretation considering a microcanonical statistical mechanical ensemble of magneto–matter homogeneous tubes and their dimensions.

| Parameter | Value |
|---|---|
| Assumed specific heat index $C_v$ | $3kT$ |
| Rounded adiabatic polytropic index $\gamma$ | 1/6 |
| Rounded e-gas work coupling constant $\eta$ | 5/4 |
| Rounded ensemble #of magneto-matter tubes | $10^{13}$ |
| Estimated magneto-matter tubes mean diameter length $l$ | 20 km |
| Estimated magneto-matter tubes mean cross section $\pi(l/2)^2$ | 314.16 $km^2$ |
| Estimated magneto-matter tubes mean volume $\pi l^3$ | 25,132.74 $km^3$ |
| Estimated mean no.moles per magneto - matter tubes | $947221 \equiv 10^6$ |
| Ideal gas-constant R | $8.3144598\ (48)\ J mol^{-1} K^{-1}$ |
| Gravitational constant G | $6.67408(31) x 10^{-11} m^3 kg^{-1} s^{-2}$ |
| Expansion scaling value $¢ = l/R_{\odot}$ | 2/7 $x\ 10^{-4}$ (see Appendix 7) |
| $v_s$ (acoustic, non diffractive mode) | $2x10^3 km/s$ |

### 4.3 From the single electron to an ensemble of *magneto-matter in thermal equilibrium*.

Consider the single electron Hamiltonian (see e.g. Eq. 89, sect. 41 in Dirac, 1967)

$$H_e = \tfrac{1}{2}\, p^2/m_e + G\, m_e M_{Sun} [r-\langle r\rangle]/\langle r\rangle^2 + \mathbf{B}(r)\cdot \mathbf{m}_e \qquad (13)$$

where $\mathbf{B} = \mu\ \mathbf{H}$, and here the magnetic permeability $\mu$ is a property of the medium to be determined.

Equation 13 provides us with a synthetic explanation for the possibility that the matter of the Sun Corona in the heights described do not cool. Equation (13) shows that when the electron is higher in the Corona, i.e. $[r-\langle r\rangle]>0$, it gains potential energy, which will be equal to the kinetic energy lost in the absence of a magnetic field or other force-field. Now, considering the correct interpretation of the CODET model in the magnetic field **B**(*r*) in the Corona, as well as our well supported understanding of its decrease with increasing *r,* the effect may result in a net increase of kinetic energy by the electron in Equation (13).

Hence, we assume that the gravitational and magnetic fields are such that a net increase in kinetic energy of the electron takes place. But we also assume that in the region of the quiescent Corona of the solar minimum there is a global thermodynamic equilibrium in the region of interest. I.e. we assume that magnetic field intensity **H** is added continually by some mechanism below this region in the Corona of our interest. This part of the Corona temperature stays constant (see Figure 3), within the resolution of our observations, possibly because of losses mainly due to irradiation (by the matter in the Corona moving possible up), which energy would be replenished with the influx of new **B**-field flux and matter from below.

Hill, 1960, in his book on statistical mechanics describes in its Chapter 7, Section 1, for a 2D absorption theory, by Langmuir, of a gas by a solid the same conditions we envision in this magneto-matter state. This is a state in which matter has coalesced completely with the magnetic field, whereas when a gas in the system is present, as described in the outline below, in which we use Berdichevsky and Schefers, 2015 to assume:

1st, that at each height layer corresponds a same value *β.* This is what Equations 9 imply. For this assumption to hold is simpler to suppose that most matter is ionized, and we can assume this plasma is neutral, with most of it frozen to the dominant magnetic field. This is supposed despite we are aware that the plasma in the corona could possess $\beta \sim 1$, although $\gamma << 1$, *see comment* toward the end of Section 4.2.

2nd, that there is in the solar corona an ideal gas of electrons, as is common in many circumstances, even as it is the case when dealing with an electron gas in the conduction band of a metallic solid, see e.g., Ziman, 1964.

3rd, that the equation of state of the gas of electrons represented by Equation 12 results from a system that is in thermal equilibrium. And because of the considered dimensions and assumptions we take into account three relevant works to which this electron gas may be subjected,

3.a, e-gas work *(i.e.* $P_e dV$*)*

3.b, gravitational work which the gas of electrons perform, and

3.c, magnetization work, like the one postulated in ‘magneto matter’ in Berdichevsky and Schefers, 2015.

The 1st through 3rd conditions explain the validity of the relationship derived in Rodriguez-Gomez, 2017. However, they also beg the question as to what is this state of matter. Same as we did in Berdichevsky and Schefers, 2015, for our study of the properties of the medium (magnetized – matter in a FR studied in-situ at 1 AU). We propose we are dealing with a 3-D Langmuir amorphous lattice as Langmuir proposed to explain adsorption of a gas by a surface of a solid, see e.g. Langmuir, 1916, 1932, Langmuir and Taylor, 1932.

At 1 AU we refer to the interpretation of observations which are well documented, see e.g. Osherovich et al, 1993, 1997, 1998, Farrugia et al, 1995, Sittler and Burlaga, 1998. They show how at different distances from the Sun there is the occurrence of a similar anomalous behavior of the electron gas, i.e. $\gamma < 1$ for an electron gas (see above mentioned thermodynamics interpretation by Berdichevsky and Schefer, 2015).

From the observations we further assume that once in the Sun corona region here discussed we are in the presence of an ensemble of uniformly magnetized matter in which '*the unit element*' of this ensemble is an homogeneous tube of magnetized matter with a current that flows along the magnetic field contributing to the generation of magnetization in the e-gas. Also we assume *'no'* matter flow, but with the exception of the current carrier(s) as postulated in the '*1st*' of '*3*' assumptions made above in this subsection 4.3. (A simple estimate in Appendix 6 gives ~ $10^{13}$ *'micro'* magnetic-matter tubes with average width/length of 20/80 in units of km in the Sun's corona region from 1.16 to 1.23 solar radius $R_{\odot}$. The ensemble number of the homogeneous tubes increases to $10^{15}$ when we consider the corona volume from 1.13 to 1.30 $R_{\odot}$.)

Notice that the physics of our approach enables vibration of the: above defined *'micro'* magneto matter tubes '*due to the breathing mode*,' which would enable thermal conduction. In this way a thermal bath for the ensemble is assumed of the above defined *tubes-ensemble* of magneto-matter. Hence, we consider that it is possible to generate the global thermodynamic equilibrium, which we interpret to be present in this featureless Sun's corona (the quiescent sun corona at solar minimum).

Thus, in our approach we assume that the quiescent Sun Corona corresponds to a time-interval where it may be considered to be composed of by an ensemble of homogeneous domains with gradual expanded extension as we move higher in altitude between 1.1 and 1.3 $R_{\odot}$ in thermal equilibrium, following the scaling law

$$l_i = ¢(r_i - r_b) + l_b$$

where we assume $r_b = 20\ km$ and $¢$ is a parameter to be adjusted, but which will allow an expansion of $Ol$ or less in the altitude interval of consideration in this study (see Appendix 7). Our 'ensemble-unit' then is the mean unit of volume of magnetized-matter in the corona. This mean unit is assumed to be magnetized matter in a tube of length $4l_i$ with magnetic field **B** along the tube axis, as well as with current density **J** (// **B**), and with cross-section $\pi l_i^2$. The index '*i*' stands for each of the layers in altitude in the Sun Corona considered in this study. It is assumed that there is no matter flow between these tubes of magnetized matter, which are considered to populate the quiescent Sun Corona in the region of study (except perhaps only for the free flow of the current carriers). The stated assumption allows us to proceed with the two possible thermodynamic conditions proposed as physical understandings of the accurate description of the model of temperature and density of electrons in the low K-Corona introduced in Section 2, see also Rodríguez Gómez, 2017. Considered geometric values are listed in Table 2.

**a- Single phase condition**

The consideration for having the 1.1 to 1.3 $R_{\odot}$ part of the corona interpreted fundamentally as a 'micro-canonical' ensemble (see e.g. 1st Chapter in Hill, 1964), i.e. a close ensemble of the magnetic tube region, allows us to test the nature of magnetic fields *anker* at both ends in the solar corona as it is assumed happens in scales of large transients, see e.g. Burlaga, 1995,

Marubachi, 1997. For information about the frame of the evolution of a magnetic flux-rope see also Berdichevsky, 2013, Berdichevsky, Lepping and Farrugia, 2003. This appears to be in consistent agreement with the models evaluation of the magnetic field **B** in this region of the 'low' Sun Corona, see Table 2.

When there is not mixing between the frozen in magnetic field electrons, i.e. the e-lattice, and the gaseous electrons, following 3.a, 3.b and 3.c it is possible to write for the e-gas

$$PV = N_{mol}RT \quad (14.a)$$

We know that the electron gas. $N_{mol}$ is equal to the number of moles, in a volume $V$, and we also know that one mole contains 6.022 x$10^{23}$ particles, which was defined by Avogadro for atoms, see e.g. Zemansky, 1957, see Table 3. *R =8.314 462 Joules K/moles* is the Reynolds constant, and $T_e$ the temperature given by the model, see Table 1.

For the gravitation field we consider just changes in altitude measured in distance from

$<r> = 1.19\ R_\odot$ which gives

$$W_{Gravitation} = GM_{sun}N_e V m_e (r - <r>)/<r>^2 \quad (14.b)$$

and on the right panel (Figure 15) for

$$W_{magnetization} = \mu \mathbf{H} \cdot \mathbf{m} \quad (14.c)$$

For the thermodynamic equilibrium in a bath of heat of the system, we use the isothermal branch for the expression describing the evolution of one mole in going from location 1 to location 2, see left panel Figure 13. In consequence, we can write the change in the internal energy of the gas, see e.g. Zemansky, 1957, for a differential change in altitude in the Sun Corona

$$dU = \delta Q - PdV + G\, M_{Sun}\, m_e/<r>^2 (r - <r>) + \mu\, \mathbf{H} \cdot d\mathbf{m}. \quad (15)$$

Under the simple consideration that quiescent conditions of the solar minimum corona implies heat equilibrium ($\delta Q = 0$) between constantly added magnetic energy from the Sun interior equilibrated by the Sun Corona radiative/convective losses, and conservation of temperature (i.e. $dU = 0$) for a dilute ideal e-gas we obtain

$T_e = T_0$ (constant, isothermal consideration) (16.a)

$N_e = N_0(<r>/r)^2$. (16.b)

The chosen distance $<r>$, from the Sun-center, is convenient for using in the figures to help compare in Table 3 the model predictions with the used theory, where $r_0 = <r> = 1.19\ R_\odot$.

**Table 3**, results for October 1 – 5, 2008, near extended solar minimum. The ± sign separates *mean value* from *variability* in interval used to compare model to theory (thermodynamic interpretation).

| **Height ($r/R_\odot$)** | 1.16 | 1.19 | 1.23 |
|---|---|---|---|
| **∍Magnetic Field ($B_{model}(r)$)·** | 0.632 ± 0.005 | 0.592 ± 0.005 | 0.500 ± 0.005 |
| **Magnetic Field ($B_{theory}(r)$)·** | 0.768 | *0.730 | 0.683 |
| **∍Model electrons *T*** | 1.665 x $10^6$ K | 1.665 x $10^6$ K | 1.665 x $10^6$ K |
| **Theory electrons *T*** | 1.665 x $10^6$ K | *1.665 x $10^6$ K | 1.665 x $10^6$ K |
| **∍Model electrons $N_e$ (nro/cm$^3$)** | 4.3 ± 0.2 x$10^7$ | 3.6 ± 0.2 x $10^7$ | 3.2 ± 0.2 x $10^7$ |
| **Theory (a) $N_e$** | 3.79 x $10^7$ | *3.60 x $10^7$ | 3.35 x $10^7$ |
| **Theory (b) $N_e$** | 4.30 x $10^7$ | *3.60 x $10^7$ | 3.20 x $10^7$ |

*Anchor value

· Model B-pressure in Gauss, about 1.233 times values from Figure 5

∍From Figures 2 and 3 and 5 respectively

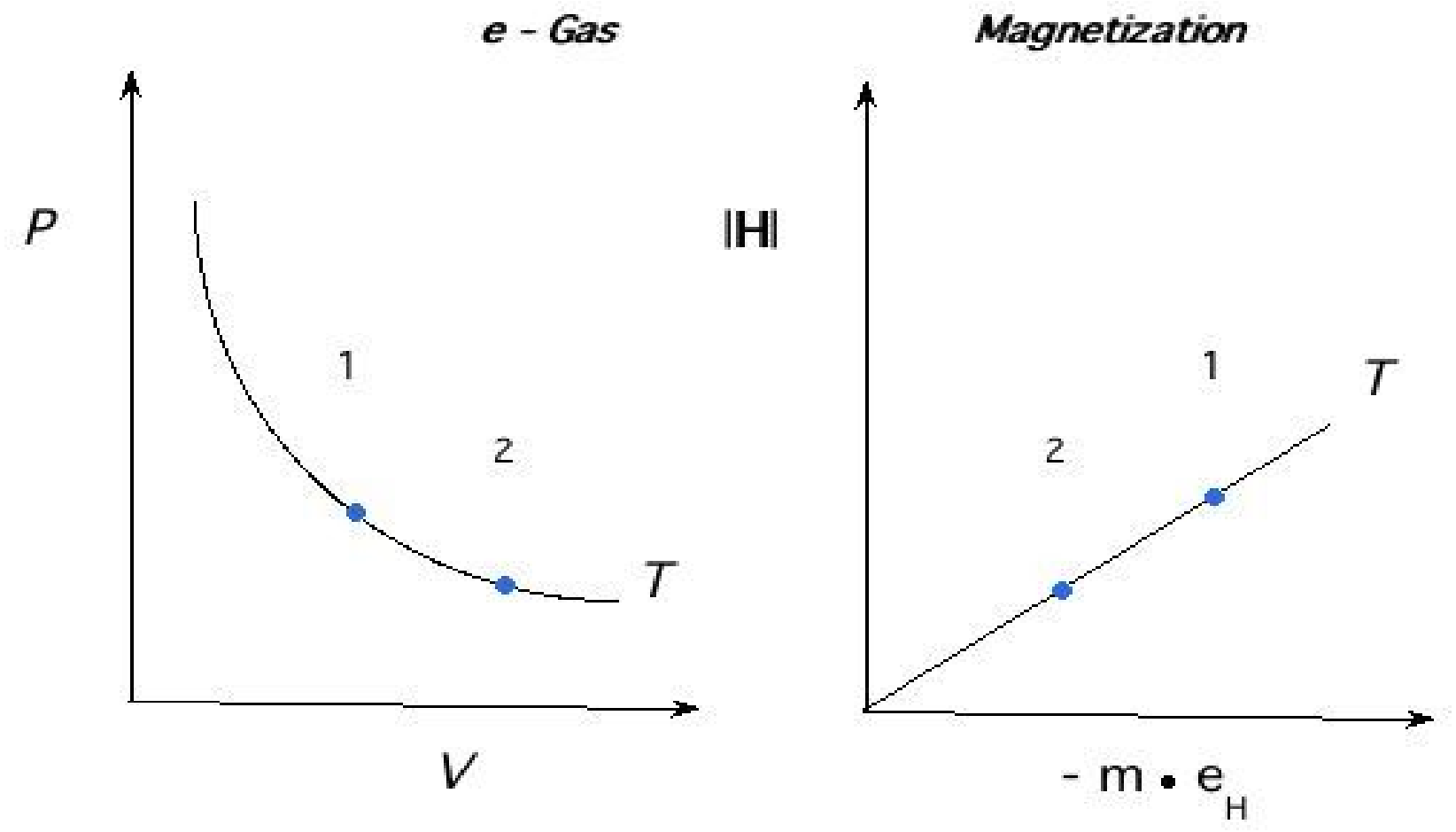

**Figure 13**. The isothermal branch of thermodynamic equilibrium in a bath of heat of the system. Location 1 at r, and 2 at the incremental higher altitude *'r+δr.'*

In this case we can make an indirect estimate of the permeability of the model when we use relationship

$$V_A(\mu)/V_A(\mu_0) = (\mu_0/\mu)^{1/2}$$

compare the estimated alfvenic speed at solar minimum at 1.19$R_\odot$ using prediction of the Alfvén from the estimated density, temperature and magnetic field estimates, and in this way we obtain $V_A(\mu_0) = 20 \pm 6\ kms^{-1}$.

We assume that that medium will differ little from the conditions deeper near 1.08$R_\odot$ in the corona from which there exist good determination of EIT wave speeds. Alfvén speed of EIT waves, here values are taken from Table 1 in Mann, et al 1999, when as in this work it is assumed that they constitute the non-compressible MHD mode (waves of the Alfvén-type) exited by the expulsion of the CME observed, i.e.

$$V_A = < V_{EIT} > = (\Sigma_i V_{EIT}(i)/n_{EIT})$$

With $n_{EIT} = 16$ (a number of 16 EIT waves associated to CMEs), in this case $< V_{EIT} > = 207 \pm 18$ $kms^{-1}$ is obtained. Therefore, we obtain the estimated permeability $\mu = 0.010 \pm 0.008\ \mu_0$, i.e. gives us a value of $\beta = 0.25 \pm 0.20$, smaller than one, surprisingly in qualitative agreement with a case study at 1 AU in Berdichevsky and Schefer.

While Table 3 shows that this result is reasonably in agreement with the model prediction, it is possible to speculate on a slightly less simple condition that would take care of the minor disagreements in *N* shown in Table 2 between CODET model and Equation 16a and b.

**b- Two phases condition**

The one phase condition's description represents the corona electrons in possibly its simplest realization. A one degree of complexity can be added. It is possible to suppose that in the corona region between 1.1 and 1.3 $R_\odot$ there exists a thermodynamic interface between the 'valence-'electrons and the 'conduction-'electrons, where the conduction electrons constitutes the e-gas. This condition of thermodynamic equilibrium of two electron phase states: the gas state increases with the volume as it occurs in the corona when transitioning from a layer at a lower altitude to one at a higher altitude. An analogy can be made with the $H_2O$ ice-water-vapor transition at T = 275.16 K, when the system isolated undergoes a volume increase in the laboratory by a controlled quasi-stationary displacement of a piston, thereby increasing the volume of the thermally isolated container of the ice and its water vapor.

Closely related to the above discussed condition of energy conservation, Equation 15, is the condition of work equilibrium. This is the expression of the Helmholtz free energy *h,* i.e.

$$d(h + PV + W_{Gravitation} + W_{magnetization}) = 0 \qquad (17)$$

and considering that

$$h = \mu_{lattice} N_{lattice} + \mu_{gas} N_{gas} = Constant \qquad (18)$$

*i*.e., we assume next that the thermodynamic phases (e-lattice and e-gas) are in equilibrium, where $\mu_{lattice}$, and $\mu_{gas}$, constitute the chemical potential of each 'e-state,' i.e.

$$VdP + dW_{Gravitation} + dW_{magnetization} = 0 \qquad (19)$$

Once more, much as we did before for a single state case we may proceed similarly with *the* two-phase state. Hence, for the condition that we attempt to explain, the global Sun Corona model outlined in Section 2 in a region between 1.1 and 1.3 $r/R_\odot$, the two phase condition in the particle density of Equation 18 allows us to write as a function of r the population for a dilute gas of electrons, the work equilibrium in the same form (Equation 14.a)

$$P\, V = N_{mol}\, R\, T_e$$

In other words. for the thermodynamic equilibrium in a bath of heat of the system we use the isothermal branch for the expression describing the evolution of one mole in going from location 1 to location 2, see left panel in the Figure 15.

For the gravitation field we consider just changes in altitude measured in distance *r* from the sun, but centered at $\langle r \rangle = 1.2\ R_\odot$, which gives Equation 14.b

$$W_{Gravitation} = G\, M_{Sun}\, N_e V\, m_e\, (r - \langle r \rangle)/\langle r \rangle^2$$

and on right panel (Figure 15) for the magnetic work (Equation 14.c)

$$W_{Magnetization} = \mu\, \mathbf{H} \cdot \mathbf{m}$$

In Equation 14.c we have the 'intensity' thermodynamic variable **H** and the extensive variable **m**, which can be written

$$\mathbf{m} = \mathbf{M}\ \Lambda \qquad (20)$$

where $\Lambda$ constitutes the cross-section –perpendicular to the orientation of the dipole vector **m** *is* relevant to the problem. In our case, at a layer located at anyone of the three altitude levels $r_i$ in the Sun Corona considered. Then at $r_b$ the magnetic flux-tube cross section is $1/4\pi l_b^2$ (see Table 2). The magnetization **M**, a constitutive property of each material –when matter-homogeneity dominates– relates to **H**, and **B** through

$$\mathbf{H} = \mathbf{B}/\mu_0 - \mathbf{M} \qquad (21.a)$$

and

$$\mathbf{B}/\mu = \mathbf{H} \qquad (21.b)$$

(rationalized MKS system of units) see e.g. Jackson, 1963. We can write the differential work contribution as the magneto-matter displacement as a differential in height (*dr*) as

$$R\,T\,dP/P + G\,M_{Sun}\,m_e\,/r^2\,dr - \mu\,\mathbf{H}(r)\cdot\mathbf{m}(r)\,dr = 0 \qquad (22)$$

which gives for the integral of gravitational work

$$W_{\mathbf{g}}\,(r - r_b) = \int_{r_b}^{r} G\,M_{Sun}\,\frac{N_e\,V\,m_e}{r^2}\,dr \qquad (23)$$

from $r_b = 1.16\ R_\odot$ to r ($r \le 1.23\,R_\odot$).

In the case of magnetization work (Equation 14.c) we consider

$$\mathbf{M} = \alpha'_{\mathbf{m}}\,\mathbf{H} \qquad (24.a)$$

i.e.

$$\mathbf{m} = \Lambda\alpha'_{\mathbf{m}}\,\mathbf{H} \qquad (24.b)$$

We use the expressions for **H** and **m** from Equations 24 in the magnetic work expression (Equation 14.c), and also use for the magnetic work the radial dependence suggested on the right panel in Figure 15. At the same time, as noticed above in the section, i.e. Equations 24, we consider that each magnetic flux tube constituting one element of the ensemble of magneto-matter in the quiescent Sun corona, each having an inverse quadratic dependence with the distance of the magnetic flux-rope (flux-tube with twist) observed in-situ, see Berdichevsky, 2013.

Hence, the magnetic work can be written as

$$W_{\mathbf{m}}\,(r - r_b) = \int_{r_b}^{r} \mu\,\mathbf{H}(r)\cdot\mathbf{m}(r)\,dr = \int_{r_b}^{r} \mu\,\alpha'_{\mathbf{m}}\Lambda|\mathbf{H}(r_b)|^2\,dr/r^4 \qquad (25)$$

when using the view of the representative magnetic field of the corona for one element in the ensemble of closed tubes of mean length $4l$ at layer b, while changing with increasing distance in a simple proportionality law $[\text{¢}(r - r_b) + l)$. In this way combining Equations 24 into Equation 25 we obtain the integration of Equation 25.

If the model provided gives isothermal conditions in the region of interest, see e.g. Figure 3, we can further write the pressure evolution observed in the region of the Corona as

$$P(r - r_b) = P(r_b)\,e^{-[W\mathbf{g}\,(r - rb) + W\mathbf{m}\,(r - rb)]/RT} \qquad (26.a)$$

where the expression (25) when using an ideal e-gas relation, Equation 14.a, gives

$$N(r) = N(\langle r\rangle)\,[r/\langle r\rangle]^2\,e^{-[W\mathbf{g}\,(r - rb) + W\mathbf{m}\,(r - rb)]/RT} \qquad (26.b)$$

Equation 26.b where a decrease in density with height as volume increase, adds the contribution of work, i.e. the explicit role of the two-phase condition through the argument of the exponential function:

$$W_{\mathbf{g}}\,(r - r_b) = GM_{Sun} N_e V\, m_e\, (r_b^{-1} - r^{-1}) \qquad (27.a)$$

as obtained from the integration of Equation 23, and

$$W_{\mathbf{m}}\,(r - r_b) = \mu\alpha'_{\mathbf{m}}\Lambda/3\,|\mathbf{H}(r_b)|^2\, l^4\,(\,[¢(r - r_b) + l\,]^3 - l^3\,)/\{l^3\,[¢(r - r_b) + l\,]^3\} \qquad (27.b)$$

As obtained from the integration of Equation 25. Equation 27.b is valid for the *mean* size element of the ensemble in this two phase approach of Sun K-Corona e-gas number density of the model for the three layers of the application of this model (the bottom one at $r_b$ = 1.16$R_\odot$, the approximately mean one at $r_m = 1.19 R_\odot$ , and the top one at $r_t = 2.23 R_\odot$), i.e. we write

$$\ln([N_b/r_b^2]/[N/r^2])\, N_{mol}RT = -\{GM_{Sun}\, N_e V\, m_e\, (r_b^{-1} - r^{-1}) + \tfrac{1}{3}\mu\alpha'_{\mathbf{m}}\Lambda|\mathbf{H}(r_b)|^2\, l\, ¢(r-r_b)([¢(r-r_b)+3/2\, l\,]^2+3[l/2]^2/[¢(r-r_b)+l\,]^3)\} \qquad (28)$$

with sub-index $i = b, m, t,$ we replaced the variable $r$ with $r_i$. where $¢ \cong 2x10^{-5}$ for $¢(r_t\text{-}r_b) \sim l$, see Appendix 7, henceforth Table 4, with the help of the values for $N_i$, the gravitational work, and $|\mathbf{B}_i^2|$ allows the evaluation of the right hand-side of Equation 28 for the difference between layers *b, and m, as well as b and t*. The difference values are listed in Table 3, and in this case explicit consideration is given to the contribution of the gravitational potential.

In Equations 28 the unknowns are $|\mathbf{H}(r_b)|$ and the cross-section *time* magnetization coupling constant ($\Lambda\alpha'_{\mathbf{m}}$). Using the constitutive relationship of the right in Equation 21 we obtain

$$\Lambda\alpha'_m \cong -2.1\text{x}10^6 N_{mol}/\text{mol Joules}/\{\tfrac{1}{3}\mu\, |\mathbf{H}(r_m)|^2\, 0.65\, 10^{10}\text{m}^2\} \qquad (29.a)$$
$$= -0.29\, 10^{14}\mu/[\tfrac{1}{3}B^2(r_m)\, 0.65\text{x}10^{10}\text{m}^2]$$

with

$$\tfrac{1}{3}\, B^2(r_m)\, 0.65\, 10^{10}\text{m}^2 = \tfrac{1}{3}\, [0.7\, 10^{-4}\ \text{Tesla}]^2\, 0.65\, 10^{10}\text{m}^2 = 6\ (\text{s/m})^2\, [\text{Joule}/(\text{Coulomb/m})]^2$$

then

$$\tfrac{1}{3}\, B^2(r_m)\, 0.65\ \text{x}10^{10}\text{m}^2 = 6\ \text{x}10^{16}/(4\pi)\mu_0\ \textit{Joule} \qquad (30)$$

We used the model value for $T_e$ and the Sun radius $R_\odot$ in Table 1. This way we obtain

$$\Lambda\alpha'_m \cong \mu/\mu_0 \, [-0.29\ 10^{14}\ \text{Joule}/[0.5 \times 10^{16}\ \text{Joule}] \quad (31)$$

and using Equations 21, the constitutive magnetic field relationship for a material, we can write

$$\mu/\mu_0 = (1 + \Lambda\alpha'_m) \quad (32)$$

with the value $\mu = 0.94\ \mu_0$ of the magnetic permeability of this thermo-dynamic state of the magnetized matter, considered with the help of the model's predictions in Section 2 for the three layers *i under consideration,* See Table 4.

**Table 4,** Homogeneous region (HomgR) magneto-matter estimated constitutive properties for T = 1.66x10$^6$ K in Corona height layers i. work evaluation is performed for a single magnetic close tube. This mean homogeneous magnetic-flux tube of cross-section $\pi l_b^2$ and length $4l_b$. For the bottom (b) layer, is assumed to change in size as the tube moves 1st to the layer mean (m), and finally to top (t).

| ***Layer (i)*** | ***Bottom (b)*** | **~Mean (m)** | ***Top (t)*** |
|---|---|---|---|
| **Height ($r/R_\odot$)** | 1.16 | 1.19 | 1.23 |
| ***HomgR [Volume]*** | $4l_b\pi l_b^2$ | $4\pi\ (l_b + ¢[r_m - r_b])^3$ | $4\pi\ (l_b + ¢[r_t - r_b])^3$ |
| ***HomgR [Section]*** | $\pi\, l_b{}^2$ | $\pi\ (l_b + ¢[r_m - r_b])^2$ | $\pi\ (l_b + ¢[r_t - r_b])^2$ |
| ***HomgR e-number*** | $N_b\ 4l_b\ \pi l_b^2$ | $N_m\ 4\pi\ (l_b + ¢[r_m - r_b])^3$ | $N_t\ 4\pi\ (l_b + ¢[r_t - r_b])^3$ |
| ***HomgR number moles*** | $\frac{N_b 4l_b\pi l_b^2}{(6.022x10^{23})}$ | $\frac{N_m 4\pi(l_b+¢[r_m-r_n])^3}{(6.022x10^{23})}$ | $\frac{N_t 4\pi(l_b+¢[r_t-r_b])^3}{(6.022x10^{23})}$ |
| ***HomgR e-gas number*** | $N_b\ 4l_b\,\pi l_b^2$ | $N_m\ 4\pi\ (l_b + ¢[r_m - r_b])^3$ | $N_t\ 4\pi\ (l_b + ¢[r_t - r_b])^3$ |
| $ln\left(\frac{N_i/r_i^2}{N_b/r_b^2}\right)RT$ | – | – 3.3778x10$^6$ | – 6.0934x10$^6$ |
| $-W_G(r_i - r_b)$ ***[J]*** | – | – 4.040x10$^5$ | – 8.092x10$^5$ |
| $-W_M(r_i - r_b)$ ***[J] (Eq. 28)*** | – | – 2.9738x10$^6$ | – 5.284x10$^6$ |

These results are consistent with the presence of a polytropic adiabatic index smaller than 1 inferred in the model, Section 2, with parameters listed in Table 1. We notice that the properties of this quiescent magneto-plasma matter does not appears to contradict the assumptions of the CODET model use of the PFSS magnetic potential field in the region of interest with magnetic permeability $\mu_0$. In our view, nevertheless, the 'model b' emphasizes that in the region of interest takes place a self-organization of an ensemble each one of them constituted by a magneto-plasma state inside a 3-D geometric shape of a tube with a current mostly align with the magnetic field present in this environment. Also the derived nature of a diamagnetic medium is consistent with the properties found in an earlier interpretation made for quite different conditions in Berdichevsky and Schefers, 2015, which was one of strongly magnetized matter at 1 AU in solar transient interval(s) amenable to the description of the magnetic flux-rope, i.e. magnetic flux-field/tube with the Lundquist type of twist.

With the non-dispersive sound speed for the assumed magneto-matter structure, and using the information on the electron gas it is possible to obtain the non-dispersive acoustic speed of the medium, see e.g. Myers, 1990

$$v_s = [\gamma k_B T_e/m_e]^{1/2} \quad (33)$$

which gives a value of $v_s$ of approximately 2,051 $kms^{-1}$ ~ $2/3x10^{-2}c$. This appears to be in agreement with the noticed failure to observe a signal of strong shock propagation in the *1.1 – 1.3* $R_\odot$ Sun Corona, when its displacement is below ~ *1,800 – 2,200* $kms^{-1}$. On the other hand the medium would allow the formation of a slow shock with about a speed of ~ 300 $kms^{-1}$ to propagate outboard until encounters at higher altitude (perhaps $r/R_\odot$ > *1.5 – 1.8*) conditions, where the system would become **B**-field dominated, i.e. $\beta < 1$, and the Alfvén speed would reach values of ~ 1,700 $kms^{-1}$ and the conditions would be given for the low solar corona shock to vanish. This has been earlier interpret as being the case in the region, and here we follow the approach by Mann et al, 1999, see also Gopalswamy et al, 2001, authors which assume about one order of magnitude higher plasma density in a region closely below our careful re-evaluation for our value with 'Model b' of $\mu$ at $r_i$, see Equations 28 and 29 and the remote observations of $N_e$ and $T_e$ by e.g. Habbal et al, 2010, see also Sections 2 and 3. (View Appendix 8, where we followed the classical analysis of conditions of wave steepening, e.g. the tutorial on the subject of shoks by Kennel, Edmiston, and Hada, 1985, for a more up-to-date discussion through modeling see e.g. Hau and Wang, 2016.)

Regarding the temperature equilibration time we consider that a transport of an infinitesimal increment/decrement of heat ±δQ will occur at close to the acoustic speed, equation (33), see e.g. Landau and Lifshiftz, 1960. When we consider the collisional nature of heat for an ideal gas as it is considered here for the electrons, this implies for one ensemble unit of cross section $\Lambda = \pi(l/2)^2$ and length $4l$ that it will equilibrate across(along) in ~ $10^{-2}$s ($4x10^{-2}$s). For the overall region of analysis from 1.16 to 1.23 $R_\odot$ the expected time of equilibration will be about two to possibly five times 150 s when considered the non-homogeneous nature of the elements part of the ensemble magnetized tubes and the corresponding thermalization by collision among these ensemble elements, see Table 5.

**Table 5**. Magneto-matter constitutive properties for thermodynamic equilibrium, case 'b' for the time interval from Jun. 1, 2008 to Jan 1, 2009. The cross-section relation between **M** and **H** is *Λ* (*Λ* = *π* $10^8$ $m^2$).

| **Case a. considering magnetization** | **with vacuum permeability $\mu_a$** |
|---|---|
| $\Lambda\alpha'_m$ (from model and estimated $V_A$) | − 0.990 |
| $\mu_a/\mu_0$ | 0.010±0.008 |
| <\|**B**\|> | 0.703x$10^{-4}$ Tesla |
| <\|**H**\|> | 70.3 Ampere-turn/meter |
| <\|**M**\|> | − 0.696x$10^{-4}$ Tesla |
| *nB* (Magnetic Energy Density) | 0.02-Joules (0.2 $10^6$ ergies) (see Figure 5) |
| Magnetic pressure with <\|**B**\|>, and $\mu_0$ | (50±25) x $10^{-3}$ Nw/$m^2$ |
| Plasma density (<Ion mass> ~ $2\ m_p$) | 1.27 x $10^{-14}$ Kg/$m^3$ |
| Matter pressure (for <T> = 1.665 x $10^6$ K) | (3 ± 2) x$10^{-3}$ Nw/$m^2$ |
| Plasma <β> < 1 | 0.025 ± 0.020 |
| **Case b. considering magnetization** | **with diamagnetic permeability $\mu_b$** |
| *$\Lambda\alpha'_m$ (from consistent model evaluation)* | – 0.06 |
| $\mu_b/\mu_0$ | 0.94±0.01 |
| <\|**B**\|> | 0.703 Gauss |
| <\|**H**\|> | 0.745 Ampere-turn / meter |
| <\|**M**\|> (sign **M** = – sign **B**) | 0.045 Gauss |
| *nB* (Magnetic Energy Density) | 0.00021 Joules (2.1x$10^3$ ergies) |
| <**J**> (from model **B**//**J** assumption) | 2.7x$10^{-6}$ Ampere/$m^2$ |
| *Magnetic pressure (for $\mu_b/\mu_0$ = 0.94)* | 0.5±0.3 x $10^{-3}$ Newton/$m^2$ |
| Plasma <*β*> *>1* *(for $\mu/\mu_0$ = 0.94)* | 2.3 ± 0.2 |

Hence, we are able to estimate a few constitutive properties of the medium using MHD in the way of Berdichevsky and Schefers, 2015, using our value for $|\mathbf{B}(r_b)|$ which the CODET model extends from the available measurement at the Sun Photosphere, see Table 2. Here, we used once more the ideal equation of the e-gas and the magnetized-matter assumption. An incomplete listing of the constitutive properties identified for the *magneto–matter steady state* interpretation here discussed of the quiescent solar corona is given in Table 5.

Accounted errors in a few estimates listed in Table 5 are based on Sections 2 and 3 explicit estimates on the observational uncertainties reported on temperature $T$, $N_e$, $B$, as well as other assumptions made, like the Alfvenic speed in the very low corona in the case of the *model a*.

**5. About the unusual hot state of matter in the Sun corona, in ours and others perspectives.**

Our approach in the previous section touches on a simple, consistent, physically well grounded explanation for how a steady state of magneto–matter can exist in such a way that a high steady temperature in a quiescent Sun corona occurs. However, the presented interpretation (so far) does not touch on the continuous energy source needed to make such a phenomenon possible. The long – term stability observed for this thermal equilibrium in the quiescent Sun corona over a very long interval of time, like it is a solar minimum. And we considered here an even more extended quiescent state of the Sun corona, in the Solar Cycle 23 – 24 minimum, which was composed of an interval of time of nearly two years of extremely low solar activity starting in or before 2008 and lasting till almost the end of 2009 . A possible source of this energy in the form of matter and magnetic field is briefly mentioned next.

We favor a recently developed view of the supply of energy sustaining the heat we model in the low Sun corona. De Pontieu et al, 2007, suggesting that the epicules Type II may be key to the corona heating, put this mechanism forward. This is so because the epicules Type II appear to replenish steadily the Sun corona with both the magnetic field and plasma from under the Sun atmosphere/corona, and by extension, provide magnetic field and matter that is needed to sustain the energy of the system in a quasi-steady-state required for our proposed magnetization-matter state interpretation and for the fact that the Sun corona is energetically an open system with steady energy loss through radiative, convective processes. This mechanism appears promising when we consider that a by product of the CODET model quantitative description of the low quiescent solar minimum corona is that the e-gas possesses an anomalous polytropic index much lower than one (see discussion of Equation 12 in Section 4).

A different view of the Sun corona than the one *we assume in this work* for the quiescent K-Corona is proposed in Bingham et al, 2009. There the Corona close to the transition region containing the chromosphere is assumed, ‘From the soft X-ray observations the corona is seen as highly filamentary, inhomogeneous structure consisting of a complex myriad of magnetic loop-like structures, with the hottest plasmas found in regions of high magnetic field activity, which also contain the largest magnetic fields.’ It is for this environment that they develop a battery of wave-particle interaction possibilities well supported by plasma theory with the hope of having the capability of heating to extremely high temperatures a region containing the largest magnetic fields, i.e. one or more large active regions (ARs). In these ARs it is possible that the, by Bingham et al named, Shukla plasma waves may make a substantial contribution to the extreme heating observed, see e.g. dispersive shear Alfvén waves (Gekelman, 1999), their possible ponderomotive force effects (Shukla et al, 2004), and modulated polarized dispersive Alfvén waves in Bingham et al, 2009, nonlinear effects in the interaction between clusters of dispersive Alfvén waves in studies by Sundkvist et al. 2005, as well as zonal flows by kinetic Alfvén waves coupling, see Sagdeev et al, 1978a,b, Shukla 2005.

The subject of application of the Bingham et al, 2009 approach combined with the assumption of high non-collisional heat conductivity is introduced in connection to active region loop(s). While in our case –methodically discussed in Section 3*,* we observe a quiescent K-corona to which our thermodynamic assumptions of (approximate) equilibrium appear to be more adequate. Hampered by a lack of in-situ observations at the transition region of the mostly neutral solar region constituted by the photosphere/atmosphere of the Sun, the low matter density (for electrons a value of $N_e \sim 10^{14}$) has been hard to distinguish from between a variety of mechanisms for the heating of the solar atmosphere in its transition from the photosphere up to higher regions, first from the chromospheres and above it, and the TR, the much less populated, optical-thin region of the quiescent solar corona.

One of a list of interesting possible mechanism for the strong heating of the Sun's atmosphere in passing from the lower altitude region of the photosphere to the higher location of the chromosphere is proposed by Goodman, 1992, 1993. The mechanism he presents is based on a medium amenable to its description in terms of a resistive MHD model that dissipates in heat the presence of waves, causing the strong increase of the temperature in altitude between the two regions below the corona.

In the corona nano-flare heating mechanism, the proposed process for the Sun's corona gaining 1 or 2 million of degrees Kelvin would be through nano-reconnection processes continuous in space and time in a fluid state with an e-gas with normal polytropic index $\gamma = 5/3$. In this model the fundamental heating mechanism, nano-flares, is the result of (nano-)reconnections taking place approximately from the base of the Sun corona, covering a region that extends over a similar altitude range comparable to the one explored in this work for the quiescent Sun corona with the CODET model. The observations of line widths (i.e. non thermal velocities) of ions do not appear to reflect the predicted non-thermal component in the plasma due to this postulated reconnection mechanism (nano-flares), which due to the optical thin nature of the quiescent corona is solely possible to observe at boundary regions of coronal magnetic holes (and/or AR) and its lack of observational support is discussed with technical detail in Brooks and Warren, 2016.

With respect to the mechanism for the heating of the Sun corona, we mention here that besides the dissipation of waves energy which Bingham et al, 2009 suggested could be applied successfully in a scenario of the Sun corona, ARs, other workers in the field have proposed *also* wave generation of heating to help explain in general the corona temperature. This electromagnetic waves heating mechanism was first introduced in the 1950's by Schatzman, see e.g. Isenberg, Lee, and Hollweg, 1999, see also Isenberg and Hollweg, 1988. However, it has shown limited observational evidence as an explanation for the waves energy dissipation picture for its existence, see e.g. Kasper, Lazarus, and Gary, 2008. Nevertheless this view has been further elaborated with the addition of turbulence considerations in the process of particle acceleration. (For a description of this mechanism see e.g. Dmitruck, et al, 2003.). (Notice that turbulence, which would obscure the wave activity at the Corona basis, could be helpful in explaining the lack of detection of the intense wave activity required for the heating needed by a few millions degree corona temperature.)

The presence of alfvénic waves, of a variety of kinds, is indeed observed all over the heliosphere up to ~ 9 AU from the Sun or even further away, and constitutes a support for this mechanism of the Sun corona heating. In particular, observations in-situ reveal the fundamental alfvénic nature of the heliospheric medium, including the solar wind beginning with the in-situ observation of the Helio mission which covers from 0.28 to 1.0 AU, see e.g. Schwenn and Marsch, 1990. So far failed to materialize the signatures of such intense MHD wave activity in the base of the low Sun quiescent corona. This observational failure cause doubts on the validity of these ideas of heating. Also, the theoretically predicted non-thermal widening of the ion particle distributions predicted by the wave generated heating proposed solution is not observed in the here discussed Sun corona regions, see the detail discussion on this subject in Brooks and Warren, 2016, and references therein.

## 6. Conclusions and a few related comments

### 6.1 A few concluding remarks

1st we point out that the quantitative quality of the description of the electrons temperature and number (per $cm^3$) with the CODET model is in very good agreement with other estimates using remote observations, including those carried out during the solar total eclipse on August 1, 2008 as presented in Sections 2 and 3, see also the marginal change in temperature of the corona between 1.1 and 1.4 solar radii indicated for the evolution with altitude of the width of the forbidden (non-thermal) lines, green and red of the iron X and XIV in the work by Mierla et al, 2008.

2nd we notice in Section 3 the apparent presence of an almost total insulation of the quiescent corona of interest in this study from other structures present at the altitude discussed in the study. The implication of this is that insulation takes place as noticed and that more is needed to understand that property, in which it is apparent that the nature of the magnetic field structure necessarily plays a central role.

3rd in Section 4, consistent with the CODET model, it is straightforward to find well funded physical arguments explaining the anomalous nature of the polytropic index of the electron gas of the Sun K-corona. Indeed an extremely low value $\gamma << 1$, is needed, see Table 1, for a CODET model solution to be consistent with observations. As is shown in Sections 2 and 3 such optimal CODET model choice provides quantitative results for $N_e$, and $T_e$.

4th The CODET model solution gives credence to the claim that the epicules type II play a relevant role in the Sun corona heating proposed by De Pontieu et al, 2009, 2011. This is due to arguments supported by observation in this work that there is a high likelihood that the electron gas has an anomalous polytropic index smaller than 1. It is worth mentioning that this is in contradistinction to the assumption of a normal gas of electrons made by Klimchuk, 2012, who dismisses the role of the heating of the Sun corona by epicules type II.

5th Section 4 tests the thermodynamic picture of a *magneto-matter steady state* hypothesis of a Langmuir 3D amorphous lattice consistent with the interplanetary plasma theory postulated by Alfvén of the freezing of plasma to the magnetic field for low-beta plasma conditions (see e.g., Alfvén, 1942). Here, the assumption is successfully tested for the case in which the CODET model shows an anomalous polytropic index when adjusted to achieve an optimal description of the quiescent Sun corona, a dominating feature in solar cycles near the solar minima.

In Section 4 we explained how the properties for which a topology of the quiescent Sun corona is assumed to be formed by an *ensemble of self-organized magneto-matter tubes* manifest its temperature state through their collective vibration and also because of thermal equilibrium being equal to the contained e-gas observed. The e-gas observations are through the irradiance of the Sun K-corona, as reported in the CODET model in Rodríguez Gómez, 2017, also documented inRodríguez Gómez et al, 2018.

6th When considering the impact at 1 AU of the magnetization (current generated B-field for ideal MHD) model b:

(1) The conditions inferred in Section 4, with a $\beta \sim 2.3$, see Table 5, would enable the presence of the slow shock mode. It will also be consistent with the empirical inference by Webb and other authors of more favorable conditions for disconnections from the low Sun corona of slow transients, inference that appears to be confirmed in statistical studies, see e.g., *Wu and Lepping, 2015*. Also these conditions are consistent as inferred by a similar common remote observation of metric/decametric radio bursts Type II of the common occurrence of the slow shock for our region of interest, as well as their vanishing further out in the more distant Sun corona where expected conditions are of a $\beta << 1$ with the **B**-field to dominate, see e.g. Gopalswamy *et al, 2001*. Consistent with our '*model b*' for the low corona the shock would result from the near explosive lateral expansion of the ejecta as we consider in detail a few technical aspects of the natural MHD oscillation modes in Appendix 8 and references therein.

(2) is consistent also with the view that: (i) The potential magnetic field which decay with distance with the 3rd power of the ratio of the distance to the sun radius $R_\odot$ to the in situ location of observation. Its value at 1 AU practically would have disappeared (would be O $10^{-3}$ nT or less). (ii) The magnetization or better call currents generated B-field will decay much more slowly, possibly linearly, giving qualitatively the observed value at 1 AU of about 2 nT (see <**M**> value in Table 5) , which depending in the history of magnitude *B* from the corona to 1 AU may range between 0.5 and 5.0 nT. Further the model view of an ensemble of homogeneous regions in the quiescent Sun low corona would be consistent with a **B**-field highly inclined to the imaginary line connecting the Sun with an observer at 1 AU from the Sun. And the slow SW predicted by our model would show an orientation most of the time in the ecliptic plane, which is not too far of observation for the slow solar wind. Also the model is in principle consistent with the frequent encounter in the slow SW of magnetic deeps (magnetic holes in the common language used by the heliospheric community).

7[th] 'Model b' is additionally consistent with the observation of a steady state slow SW that displays the empirical evidence of the existence of contiguous regions in overall pressure balance as it can be understood assuming that MHD is a good representation of the medium, but as usually assuming a vacuum permeability ($\mu_0$). This is a consistent result, when we consider the uncertainty of the plasma and magnetic SW field measurements in empirical in-situ observations near 1 AU that makes it difficult to distinguish between the two close permeability values, μ and $\mu_0$ (see Table 5, see e.g. the identification of pressure balance for contiguous MHD domains in Burlaga, 1968).

**6.2 Comments – about/related to – the work**

The magneto-matter tubes theoretical representation we used posses a length scale $l$ = 20 km. These tubes are a mathematical/theoretical simplifying tool that appear to be well supported by extrapolation to the low corona from observations both at 1 AU (Berdichevsky and Schefers, 2015) and near the Sun as discussed in Section 3 (also, May 2019 private communication by Alzate). This ensemble analysis allows us to build the unit block on which we assumed the *magneto-matter state* of the quiescent Sun's corona exists. In this way we pursued the use of statistical mechanics to estimate a few constitutive properties of the *magneto–matter steady state* that occupy the region of the Sun's corona $1.1 < r/R_\odot < 1.3R_\odot$ region. At the center of our results is the determination of an estimated value for the diamagnetic permeability of the medium, and along with other properties listed in Table 5. These *magneto matter* properties listed in Table 5 correspond to a state of matter that satisfies the ideal–MHD conditions. Hence, the medium preserves magnetic flux, i.e. it preserves helicity, which appears to be an observationally most probable property of the medium, see e.g. Antiochos, 2013 discussion on the subject.

On the subject of the time of equilibration implied in our deduction for our *magneto–matter state*, notice that it is consistent with the ansatz made on the thermodynamic state of equilibrium of the quiescent sun corona at solar minimum when a transient breaks these quiescent conditions after a relatively short time lapse interval order of magnitude comparable to the microscopic/macroscopic equilibration times of second-fraction/minutes, with consistent observation indicating that it appears to be that in most cases the corona quickly recovers its previous existing equilibrium condition(s). The manifestation of a coronal slow forward shock appears consistent with metric Type II remote observations in the low Sun Corona in our region of interest with a plasma $\beta > 1$. On the other hand only the very fast shocks with speeds of more than 1,500 $kms^{-1}$ in the time of the solar minimum here studied in $r/R_\odot$ would propagate further from this Sun Corona region into the heliosphere. This is the result of the observations, see Section 3, and its validation of a low density of the region, dramatically close to one order magnitude less than what is estimated in earlier works, and the intensity of the magnetic field showing also a decrease, but only by a factor less than two, see earlier studies, e.g. Saito, Poland, and Munro, 1977.

As a corollary of Point 3 (Section 6.1) we propose that there is in the Sun's corona a *steady state of magneto–matter* that appears to explain well the successful CODET model quantitative reproduction of the estimated electron density and temperature in a low solar corona extending in radial distance r for $1.1 < r/R_\odot < 1.3$ for the quiescent conditions. This is our ansatz, an *amorphous 3-D Langmuir lattice* constituting a magneto – matter state in thermodynamic equilibrium (Berdichevsky and Schefers, 2015, see also e.g. Langmuir, 1932, 1934, Alfven, 1942). These ideas support the presence of magnetization-work, which in this case the CODET model enabled, providing a polytropic anomalous index $\gamma \sim 1/6$.

In our analysis it is worth seeing if we can re-interpret this perspective from a chemistry approach, see e.g. Robitaille, 2013 in which a discussion is proposed based on the extreme acidity of the medium as conveyed by spectroscopic remote observation, see in this regard the pioneering, transformative work on the field of chemistry by Pauling, 1960 (and for a physics perspective of a solid state of matter Ziman, 1960). There, in a series of works, Robitaille argues that the plain gas interpretation can not hold, and coalescence of a particular kind is attributed to be present, which is viewed as likely requiring extreme low temperatures in the realm of a few to a fraction of degrees Kelvin, assuming laboratory conditions with matter for which condition(s) the magnetic field does not play a central role.

It is interesting here to point out that in this work we can agree with the freezing of the matter, most of the matter, when using the ideas of a hot magneto-matter state with a *simple* 3-D Langmuir amorphous lattice state nature (Berdichevsky and Schefers, 2015). And *maybe* our proposed anomalous e-gas in our coalescent *magneto matter state* could have correspondence with the valence electrons while the particles carrying the current could be interpreted as the ones constituting the conduction(s) band(s) of their proposed unusual, extremely high acidity chemical manifestations of the Sun corona optical-spectra studies as seen by chemist scholars.

With regards to the particle population of the region of the corona considered it is highly relevant to point out that only a fraction of it (less than ~ $1/15^{th}$ of the matter contained in the solar corona, estimated by the CODET model) becomes the solar wind, observed in-situ from near 0.28 AU from the Sun center and beyond, i.e. about 10 e/cm$^3$ at 1 AU. See Section 3.3, 3.4 and also 3.5. For this result we are consistent with the generalized view in the field that most of matter does not escape the quiescent solar corona. This is a subject central to the views of the geophysical space community and part of the distinction frequently made between two Solar corona regions, the one of 'open' and the other of 'closed' field lines. Here we addressed the 'closed' magnetic field region corona. Solar Probe mission's current and future observations will test these views much closer to the Sun (up to 0.1 AU or better).

Returning to the subject of the 2$^{nd}$ paragraph from the top in this Section (Section 6) it is worth considering how it becomes intriguing that there is an almost total heat insulation between the quiescent Sun corona and other structures present at contiguous places at the same height above the photosphere in the altitude range from about 1.12 to $1.25 R_\odot$. It is also worth mentioning that for the long standing quiescent intervals, beyond hours, even days if not months of observed apparent equilibrium occurred side by side with much cooler prominence/hotter AR defy understanding. Id est, they appear to corroborate that the sound derivations by Spitzer, 1956 of heat conductivity in a magnetized plasma, as well as the more detailed derivations of Braguinskii, 1965 appear to not apply. Either we observe magnetized plasma regions hermetically insulated thermally from each other or there are more cumbersome conditions. One such condition could involve some cooling mechanism which would certainly involve heat (and possibly also matter) transport, keeping regions located in close neighborhood at huge temperature differences. A circumstance that, as was pointed out by Withbroe in 1988, we still fail to understand.

Although we argue for epicules type II as a possible source of the energy keeping in 1 to 2 million degrees the *T* of the Sun quiescent corona, identifying with any certainty its source is outside the scope of this work. This is the case, as well, for several other areas here introduced which require further investigation, e.g. the estimate on the value of the chemical potentials of the valence and conduction electrons of the medium, our assumption that enables a quite satisfying evaluation of the permeability of this diamagnetic medium, see paragraphs above, and also its value listed in Table 5. The study of some/several of these unknowns will eventually be performed elsewhere.

**Acknowledgements:** D.B.B. thanks his father Carlos David for his loving support of a whole life. Further D.B.B. work is offered in the memory of his loved wife, the late Maria-Cristina Beatriz, neè del Campo y Charles. D.B.B. further wishes to thank Santiago Berdichevsky for a careful reading of the manuscript helping to simplify and make more accurate/understandable the language for the benefit of the reader. We further owe thanks to many coworkers in the division 670 of NASA/GSFC among which we here remember the help received from Karin Muglach for providing one of us with up-to-date review work on the observational understanding of the solar atmosphere as well as helping with the interaction to other colleagues/experts on the subject of the Sun's atmosphere. I further am very thankful to the patience shown by Peter Young through extensive conversations with him while I did progress in the writing of this work. Generous conversations and suggestions I received from Richard de Vore, James Klinchuck. On the subject of solar corona irradiance I owe advice and clarifications as well as useful literature on solar corona line widths observations to Adrian Daw and Douglas Rabin. I also owe Spiro Antiochos for literature suggestions onto the subject of heat conductivity in magnetized plasmas, and to Gopalswamy on the shock propagation conditions in the low Sun corona.

## Appendix 1

With Figure A1. We assume for simplification a simply spherical geometric shape of the Sun. Here we look at it just from the perspective of its projection onto the plane of the sky. Further we use the spherical coordinate $\theta$ to distinguish the two principal solar corona regions, the ones of the polar CH at the solar minimum from the region we consider to be mainly of the quiescent corona. $\theta$ shown in the Figure A1, is the apparent limit between these two regions. This occurs approximately at $\theta \approx 30°$. This is the situation corresponding to the coronal hole at the time interval of minimal activity in the solar cycle, i.e. encompassing the solar minimum. The corresponding area of irradiance observed at the time, shown as a projection in the plane of the sky in Figure A1. amounts to the integral

$$A_{Coronal\ hole} = 1/2\, r^2 \int_0^{2\pi} d\phi \int_0^{\pi/6} d\pi$$

With two coronal hole regions symmetric to each other, the area from which the irradiance of the Sun remotely observed is $2A_{coronal\ hole}$ while the semi-sphere area is $A_{semi-sphere} = 2\pi R^2$ . Then, we consider

$$A_{semi-sphere} - 2A_{Coronal\ hole}$$

Hence, the percent of irradiance resulting from the quiescent Sun corona we consider is

$$100[A_{semi-sphere} - 2A_{Coronal\ hole}]/\ A_{semi-sphere} = 86.6\%$$

and when we consider that the irradiance from the coronal hole region appears suppressed compared to the one we use, essentially we are able to neglect that part as a contribution in the modeling of the corona irradiance as a function of the high in the low corona that is the focus of the Section 4 discussion.

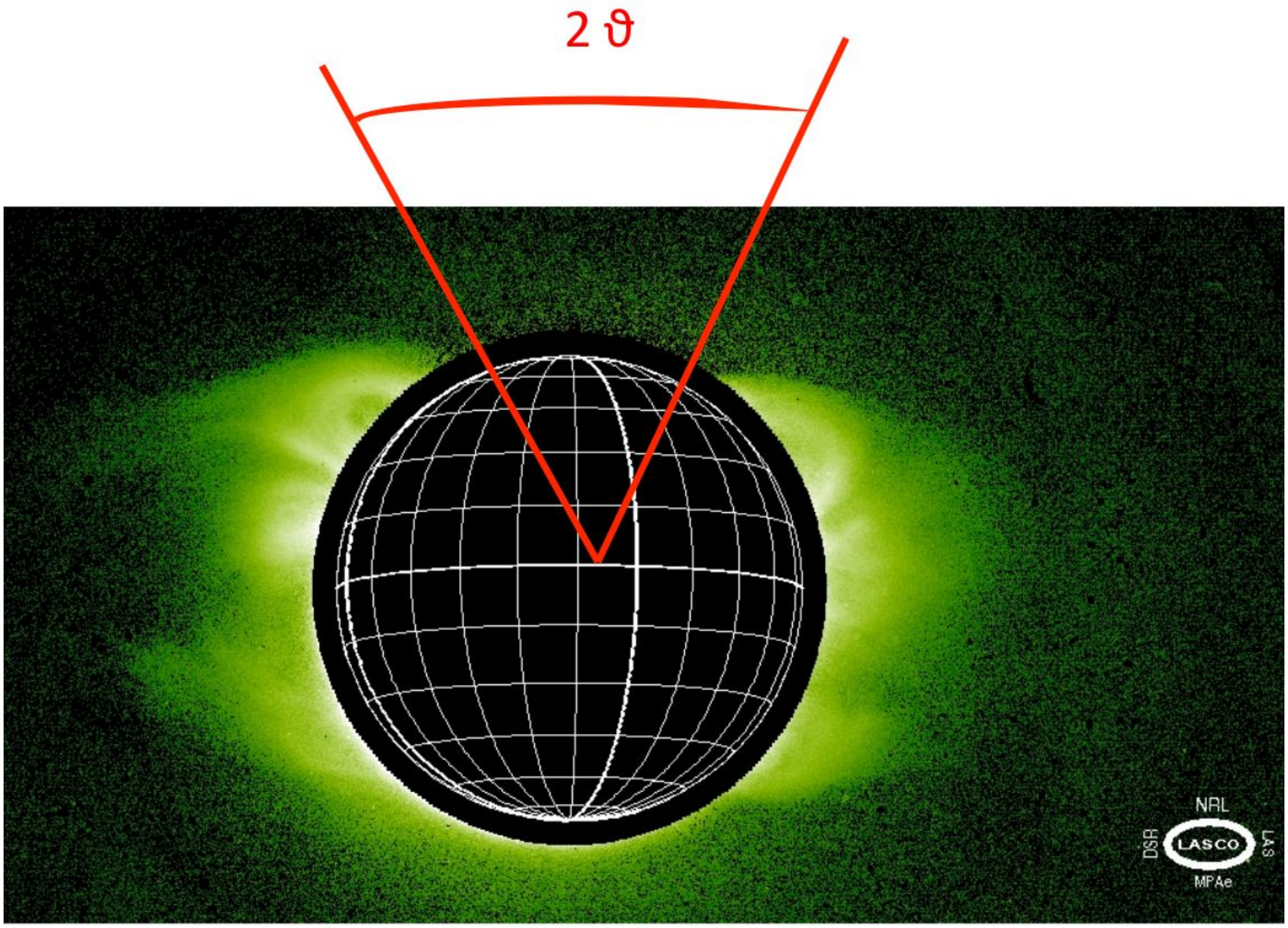

**Figure A1**. The quiescent Solar Corona in Fe XIV, the region source of the slow solar wind. Borrowed from the WWW site of LASCO/SOHO mission.

## Appendix 2

Figures 13 and 14, show SW parameters for intervals of 27 days each corresponding to April – May and July – August respectively. From top to bottom panels display $|\mathbf{V_{SW}}|$ the magnitude of the convection velocity of steady state magnetized matter, its key attributes described in Parker, 1958, 2nd from the top is the proton density. 3rd the protons kinetic energy distribution in the SW frame of reference ($V_{th}$ in the common language of Helios-physics). The 4th, 5th and 6th panel from the top respectively give the magnitude and angular orientation of frozen matter magnetic field in the SW.

In the 1st panel conventional conditions associated with the convection of the heliospheric plasma away from the Sun are marked distinguishing slow from intermediate (red line) and intermediate from fast (black) SW. The solar wind with speed above the black line is well tested to originate at a solar coronal hole, in this case an equatorial solar coronal hole. The solar wind at above 700 $kms^{-1}$ is understood to be unimpeded solar wind from a CH. Between 500 $kms^{-1}$ and 300 kms$^{-1}$ is a region resulting from a variety of coronal regions, they often include also transients (interplanetary manifestations of coronal mass ejections ICME). The nature of the solar wind from the intermediate region is not easy to connect to a specific solar region and tends to show large fluctuations in the orientation of the magnetic field, panels 6$^{th}$, 7$^{th}$, and 8$^{th}$. Before a high speed stream with origin in a CH region there is the known co-rotating interaction region (CIR) where the intermediate or slow solar wind is compressed showing an enhancement in density (2$^{nd}$ panel) and magnetic field intensity (4$^{th}$ panel). In the case of it being overtaken by an interplanetary shock the proton thermal velocity $V_{th}$ will often show a substantial increase. The alfvenic behavior of the fast speed solar wind in the 5$^{th}$ and 6$^{th}$ panels shows strongly, nearly periodic oscillation of the magnetic field while the intensity stays without variation. The azimuth angle shows the polarity of the CH region and often with extreme accuracy (away or inward from the location of the Sun). When comparing the period intervals of Figures 13 and 14 it is striking to notice in both cases the equatorial presence of CHs in the corona and in Figure 13 the near absence of very slow SW with velocities of 300 kms$^{-1}$ or less. A simpler SW organization of the in the ecliptic observed SW is seen in Figure 14 for the time interval of July – August. Figure 14 shows almost the manifestations of the SW from 'closed field regions' a denomination that applies well to the quiescent corona central to this study.

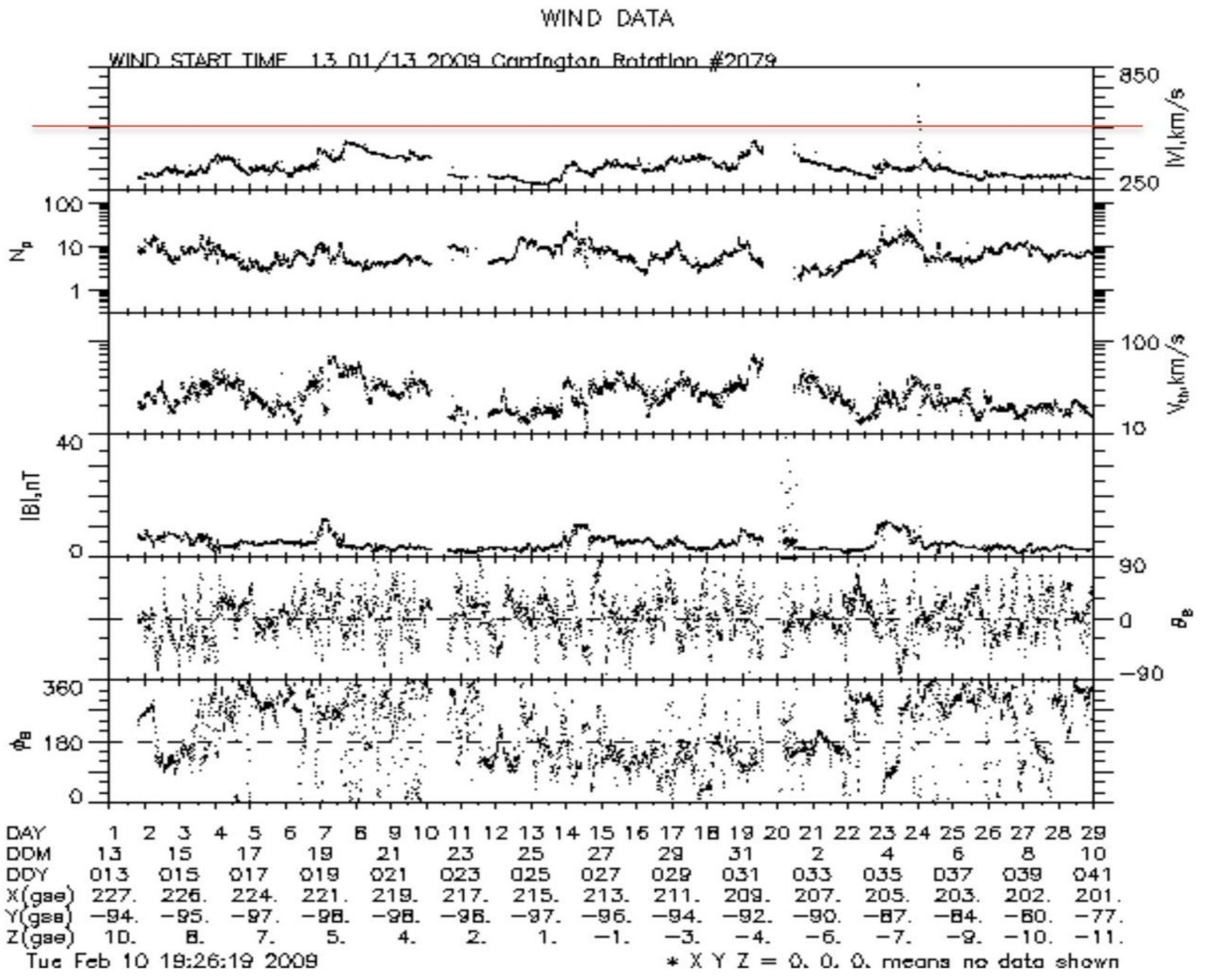

**Figure A2.a**. similar as Figures 13 and 14 but for Carrington rotation 2079, starting Jan. 13, 2009[9].

[9] https://cdaweb.gsfc.nasa.gov/cgi-bin/gif_walk
Plot type wind 27 days survey, specific date 200913

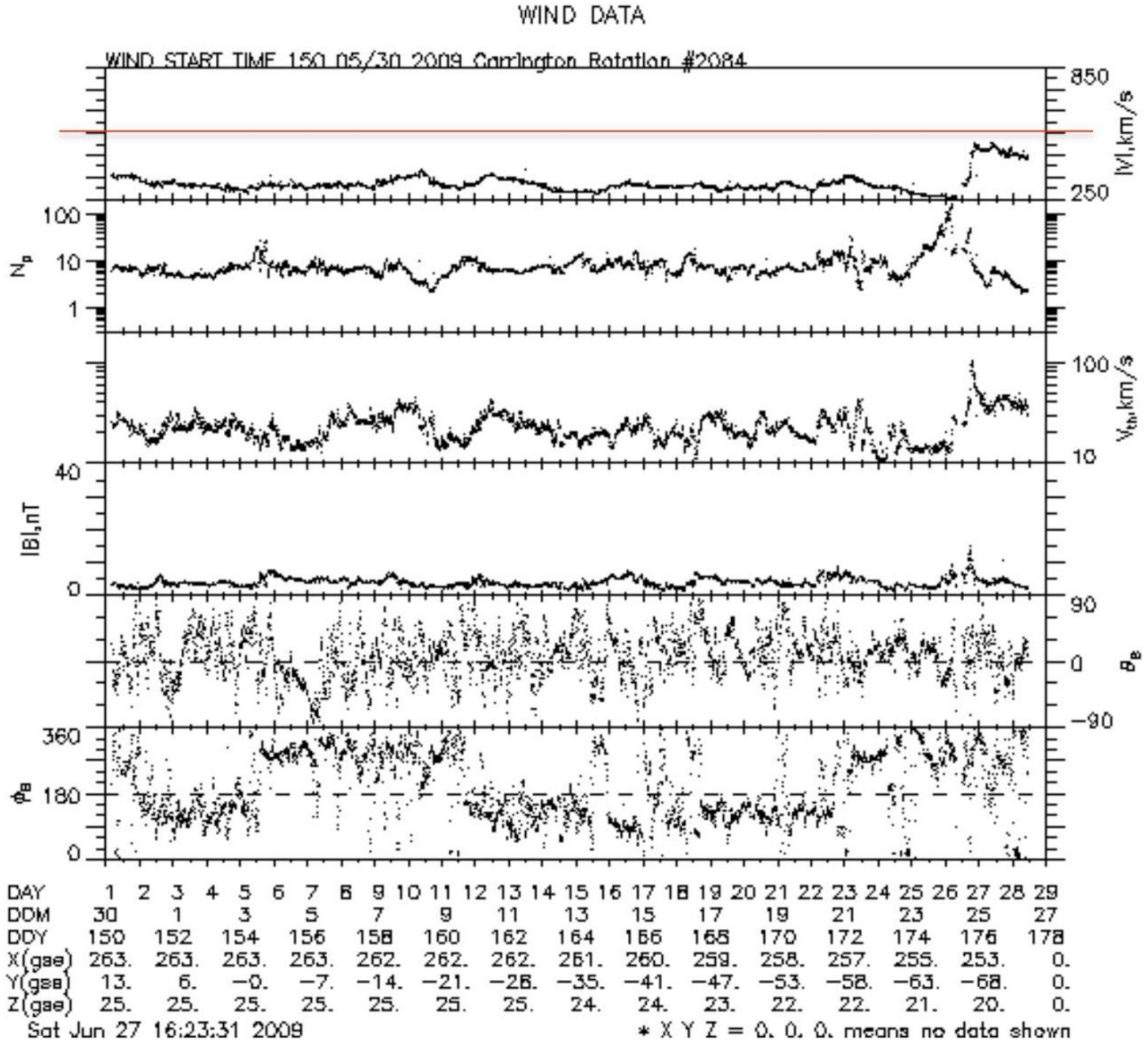

**Figure A2.b**. ibid Figure A2.a for a Carrington rotation very close to the solar minimum May. 30, 20009, Carrington rot. Number 2084[10].

Figures A2.a and A2.b again show the same organization as in Figures 13 and 14 (|$\mathbf{V}_{\mathbf{sw}}$| in the 1st panel, $N_p$ in the 2nd, $V_{th}$, in the 3rd, and the magnetic field intensity and orientation in the 4th, 5th, and 6th panels) appear to indicate a SW that although responsive to the effect of equatorial CH originated SW is at most tangentially impacted while a very slow solar wind appears to be present in the case of figure A2.a and the high speed SW almost totally absent during the time nearest to the solar minimum, which was around April 2009.

In all time intervals illustrated, both in Figures 13, and 14 as well as A2.a and A2.b, the magnetic field is substantially below long standing average, and that is true too for the number of particles per $cm^3$ indicating after Parker that we are dealing with a lesser solar output of energy into the corona, i.e. matter and magnetic field contributing to an unusual cool corona (< 2 x$10^6$ K).

[10] https://cdaweb.gsfc.nasa.gov/cgi-bin/gif_walk
Plot type wind 27 days survey, specific date 2009150

**Appendix 3**

Using similar assumptions as in Appendix 1 for the simple estimates on the geometry considered for the presence of other that quiescent solar corona region that influence the irradiance of the solar corona we are able to estimate both i) the e-number of 3x10$^7$ e/cm$^3$ at 1.23$R_\odot$ as derived via the CODET model for solar minimum consistent with the inverse tomographic determination of e-irradiance in an homogeneous K-corona in years 2008 – 2010 of minima solar activity, and ii) long time average(s) of particle number from in-situ proton measurements (Priest, 2014), which is well approximated by $<p_{1AU}>$ ~ 10/cm$^3$, in which it is relevant to consider that other ions are present with more stripped e than just $H^+$, see e.g. Berdichevsky et al, 1997.

Then it is straightforward to obtain the estimate of the e-number that contribute to the solar K-corona irradiance at r = 1.23$R_\odot$. This is given by considering the whole region which then becomes part of the equatorial streamers as is shown in Figures 8. This number is given as

$$-Number(r) \sim 0.9\ 4\pi\ r^2\ cm\ N_e$$

where 1 cm applies because the evaluation is given for number of particle per cm$^3$ both in the remote and in-situ estimates. Because we are counting over all particles streamed in the SW, using the average here is a good approximation particularly for the case of an interval during solar minimum.

The same estimate at the Lagrange point L1, for the measurement by Wind during a long time averaging extends the value to all the likely regions of 30° width of a slow solar wind heliospheric plasma sheet containing the heliosphere current sheet is reasonable. Hence, the number over this ‘disc’ of particles with the assumption of having an homogeneous outflow equivalent to 2/3 of the whole region, also a strong although reasonable approximation to make, indicates that the total streamers outflow, i.e., the solar wind in regions where in-situ observations exist (> 0.28 AU) is given by the expression

$$\text{e-Number}(r_{1AU} = 1\ \text{AU}) = 2\pi\ 20°/180°\ r_{1AU}^2\ \text{cm}\ N_{SW}$$

Then, we obtain

$$\text{e-Number}(r)\ /\ \text{e-Number}(r_{1AU} = 1\ \text{AU}) = 172\ \text{e}$$

which is about 16 times larger than the observed ~ 11.2 e in the SW with origin at the streamers, which was assumed in simplified estimates of total particles at both regions. In addition we have to consider that the inverse evaluation of the e-density from the radiance in the K-corona introduces an uncertainty that can be reasonably assumed to be of about 30%. The uncertainties in the in-situ measurement of SW particles is much smaller, in the range of 1.5 to 3 p per $cm^3$. In both cases we deal with larger uncertainties with regard to the simplified assumptions regarding the geometry both in the corona and at 1 AU.

**Appendix 4**

For the purpose of estimating the transient presence we proceed to consider at solar minimum MC, i.e. Magnetic flux-ropes, and those structures lacking –*on average*– the smooth rotation of the magnetic field and classified by Lepping and co-workers as MC-like (CL). We consider that these events will be weak, and independent of their latitude of origin, that the dominant polar coronal hole presence will direct them to be located and also mostly aligned with solar equator, while having on average an extension of in longitude of 45°, i.e. 0.28 AU, with a diameter of about 0.06 AU. Considering 1 MC and 'a' CL in 2008 we estimate statistically that the number of transients contributing to the result in 2008 to be

$$\pi\,(2 - \tfrac{1}{4})/(\pi\,\tfrac{1}{4})$$

for MCs, when adding CLs the number climbs to ~15. However, we have to distinguish the impact on the CODET modeled temperature and density of the K-corona to be different, since in their majority the CL are smaller structures than the MCs. In this way we reach a total of large de-connections from the Sun with different properties with respect to its quiescent features per day, described here as about

2 (#MC)

where we purposely try to overestimate their impact by considering that they displace slowly and expand more quickly so as to cover about ⅓ of the solar disk for ⅓ of the day, with their impact on the analysis with the CODET model per day to be about

$$2\,(\#MC/y)/2\;\; \tfrac{1}{3}\pi/(\,\tfrac{1}{2}\;\; 2\pi)\; \tfrac{1}{3}d/d = 7/365d\,(\tfrac{1}{3})^2 < 1\%$$

of the solar radiance in year 2008, when only one MC and no CL were observed, see Wu and Lepping, 2014. Our estimate is based on the same amount of MC and CL generated for that year which is larger than the specific interval of our study, which is focused on the last 2/3 part of year 2008. Their contribution would amount to a systematic error in relation to the presence of intervals, most likely of depletion in the temperature, but also affecting the e-count (density number) of electrons of about Area in units of MCs*#MC&CL*duration observation/Area Corona*total time observation.

## Appendix 5

Using Equation 11 and assuming we are in the presence of a gas of electrons that behaves in an adiabatic process as an ideal gas with a polytropic index

$$\gamma = \zeta/\alpha$$

that with the values $\zeta$ and $\alpha$ in Table 1 gives a $\gamma$ very close to 1/6, i.e. quite anomalous for being much smaller than 1. Using Eq. 4 in Berdichevsky and Schefers, 2015 the anomalous polytropic index $\gamma$ can be interpreted as relating to the occurrence of two coupled works by the e-gas, i.e. the usual *'pdV'* work associated to an ideal gas and a magnetization work '**B** ·•*d***m**'

Using the expression

$$\gamma = 1 + R/c_V - R/c_V\eta$$

in Equation 7, also in Sect. 2 of Berdichevsky and Schefers, 2015 for the e-gas, and using $\gamma = 1/6$ it is straightforward to find the coupling constant $\eta = 5/4$ value between the two works in an adiabatic process (i.e. $\eta$ = **B**·*d***m**/*pdV*), e.g. at the passage of a non-dispersive pressure wave in the assumed 3D Langmuir magneto-matter steady-state in the Sun corona. Further notice that for the considered ideal gas

$$T \propto N^{-1+\gamma}$$

an increase in density generates a cooling of the structure, which is an outcome of the $\gamma < 1$ anomalous nature of the medium.

## Appendix 6

Here we look at the already derived possible extension in volume of the quiescent solar corona between 1.15 and 1.25 $R_\odot$. Then, the volume value is

$$\sim 0.86\ 4\pi/3\ R^3\,(1.23 - 1.16)\ R^3 \approx 2.5\text{x}\ 10^{17}\ \text{km}$$

This region volume when divided by the assumed mean tube structure of

$$\pi(l/2)^2 4l = 2.5x10^{44} km^3$$

gives an ensemble of mean homogeneous regions of $10^{13}$ elements, good for statistical mechanics/thermodynamics study

## Appendix 7

Here we make the simplifying geometric assumption

$$\not{c}(r_t - r_b) = (1.23 - 1.16)\, l$$

i.e.

$$0.07\ ¢R_{\odot} = 0.07l$$

hence $¢ = l/R_{\odot}$

**Appendix 8**

As it is known, the wave modes in MHD are the fast, intermediate, and slow mode propagation, and here we follow in our notation Kennel, Edmiston, and Hada, 1985, i.e. for these MHD wave speeds we use $C_i$, *with i = Fast, Intermediate, Slow,* as well as $C_S$ = *gas speed,* and $C_A$ = *Alfven speed* respectively. Hence, we can write:

$$2\ C^2_{Fast} = \{( C_A^2 + C_S)^2 + [( C_A^2 + C_S^2)^2 - 4\ C_A^2\ C_S^2 \cos(\theta)]^{1/2}\}$$

$$C^2_{Intermediate} = C_A^2 \cos^2(\theta)$$

$$2\ C^2_{Slow} = \{( C_A^2 + C_S)^2 - [( C_A^2 + C_S^2)^2 - 4\ C_A^2\ C_S^2 \cos(\theta)]^{1/2}\}$$

where $\theta = arccos(\mathbf{B}_{up} \cdot \mathbf{n}_{Shock}\}$, i.e. $\theta$ is the angle between the direction of the magnetic field upstream of the shock ($\mathbf{B}_{up}$) with the shock normal ($\mathbf{n}_{Shock}$). Further we notice that in this environment our models gives that $C_S^2 > C_A^2$. And the Kennel, Edmiston, and Hada analyses indicate that the *Slow-Shock* steepens fastest when $C_S^2 > C_A^2$, quickly generating a shock wave. Notice we here refer to the condition in which $\theta \sim 0$, which is consistent with the geometry, **B**-field approximately parallel to the surface of the Sun, and physics of the occurrence of strong density fluctuation limit for a near parallel shock $\delta N \sim N$. I.e. our problem appears consistent with the remote metric/decametric Type II radio burst observations interpreted commonly as manifestations of shocks near the TR in the low Sun corona *only,* soon after the CME starts, as it was argued in the past, see e.g. Gopalswamy et al 2001, Mann et al, 1999.